\documentclass[journal]{IEEEtran}
\usepackage{amsmath,amsfonts}
\usepackage{algorithmic}
\usepackage{algorithm}
\usepackage{array}
\usepackage[caption=false,font=normalsize,labelfont=sf,textfont=sf]{subfig}
\usepackage{textcomp}
\usepackage{stfloats}
\usepackage{url}
\usepackage{verbatim}
\usepackage{graphicx}
\usepackage{cite}
\usepackage{acronym}
\usepackage{booktabs}
\usepackage{ifpdf}
\usepackage{url, flushend}
\usepackage{multirow} 
\ifpdf
\else
\fi

\usepackage{cite}
\usepackage{balance}
\usepackage{bm,comment,color}

\usepackage{amsmath}

\usepackage{array}
\usepackage{amsfonts} 
\usepackage{amssymb}
\usepackage{esint} 
\usepackage{units}
\usepackage{multirow}
\usepackage{comment}

\usepackage{color}

\definecolor{AlirezaPurple}{RGB}{150, 0, 250}

\usepackage{pgfplots}
\usepackage{tikz}
\usetikzlibrary{calc}
\makeatletter
\newcommand{\gettikzxy}[3]{%
  \tikz@scan@one@point\pgfutil@firstofone#1\relax
  \edef#2{\the\pgf@x}%
  \edef#3{\the\pgf@y}%
}
\usetikzlibrary{spy,backgrounds}
\pgfplotsset{compat=newest}
\usetikzlibrary{plotmarks}
\usetikzlibrary{arrows.meta}
\usepgfplotslibrary{patchplots}
\usepackage{grffile}
\newlength\fheight 
\newlength\fwidth 
\usepgfplotslibrary{fillbetween}

\acrodef{6g}[6G]{sixth generation}
\acrodef{ael}[AEL]{angle error loss}
\acrodef{ad}[AD]{autonomous drive}
\acrodef{adas}[ADAS]{advanced driver assistance system}
\acrodef{ai}[AI]{artificial intelligence}
\acrodef{ao}[AO]{alternating optimization}
\acrodef{aoa}[AOA]{angles-of-arrival}
\acrodef{aod}[AOD]{angles-of-departure}\acrodef{bs}[BS]{base station}
\acrodef{cdf}[CDF]{cumulative distribution function}
\acrodef{crb}[CRB]{Cram\'er–Rao bound}
\acrodef{ccrb}[CCRB]{constrained Cram\'er–Rao bound}
\acrodef{dbscan}[DBSCAN]{density-based spatial clustering of applications with noise}
\acrodef{esprit}[ESPRIT]{estimation of signal parameters via rotational invariant techniques}
\acrodef{fim}[FIM]{Fisher information matrix}
\acrodef{gd}[GD]{gradient descent}
\acrodef{gnss}[GNSS]{global navigation satellite system}
\acrodef{gps}[GPS]{global positioning system}
\acrodef{hwi}[HWI]{hardware impairment}
\acrodef{imu}[IMU]{inertial measurement unit}
\acrodef{ip}[IP]{incidence point}
\acrodef{iqi}[IQI]{in-phase and quadrature imbalance}
\acrodef{isac}[ISAC]{integrated sensing and communication}
\acrodef{las}[L\&S]{localization and sensing}
\acrodef{los}[LOS]{line-of-sight}
\acrodef{ls}[LS]{least squares}
\acrodef{mae}[MAE]{mean absolute value}
\acrodef{map}[MAP]{maximum a posteriori}
\acrodef{mc}[MC]{mutual coupling}
\acrodef{mcrb}[MCRB]{misspecified Cram\'er–Rao bound}
\acrodef{mimo}[MIMO]{multiple-input multiple-output}
\acrodef{mle}[MLE]{maximum likelihood estimator}
\acrodef{mlb}[MLB]{mismatched lower bound}
\acrodef{mmwave}[mmWave]{millimeter wave}
\acrodef{mmse}[MMSE]{minimum mean square error}
\acrodef{mpc}[MPC]{multipath component}
\acrodef{nlos}[NLOS]{non-line-of-sight}
\acrodef{ofdm}[OFDM]{orthogonal frequency division multiplexing}
\acrodef{ota}[OTA]{over-the-air}
\acrodef{pan}[PAN]{power amplifier nonlinearity}
\acrodef{pdf}[PDF]{probability density function}
\acrodef{pn}[PN]{phase noise}
\acrodef{prs}[PRS]{positioning reference signal}
\acrodef{pss}[PSS]{primary synchronization signal}
\acrodef{rel}[REL]{response error loss}
\acrodef{rev}[REV]{rotating element electric field vector}
\acrodef{rf}[RF]{radio frequency}
\acrodef{ris}[RIS]{reconfigurable intelligent surface}
\acrodef{rss}[RSS]{received signal strength}
\acrodef{rtk}[RTK]{real-time kinematic}
\acrodef{rtt}[RTT]{round-trip-time}
\acrodef{slam}[SLAM]{simultaneous localization and mapping}
\acrodef{snr}[SNR]{signal-to-noise ratio}
\acrodef{ssb}[SSB]{synchronization signal/physical broadcast channel block}
\acrodef{tdoa}[TDOA]{time-difference-of-arrival}
\acrodef{toa}[TOA]{time-of-arrival}
\acrodef{ue}[UE]{user equipment}
\acrodef{ula}[ULA]{uniform linear array}
\acrodef{ura}[URA]{uniform rectangular array}
\acrodef{va}[VA]{virtual anchor}

\long\def\comment#1{}

\DeclareMathOperator*{\argmax}{arg\,max}
\DeclareMathOperator*{\argmin}{arg\,min}

\newfont{\bbb}{msbm10 scaled 700}

\newfont{\bb}{msbm10 scaled 1100}
\newcommand{\CC}{\mbox{\bb C}}

\newcommand{\av}{{\bf a}}
\newcommand{\bv}{{\bf b}}

\newcommand{\dv}{{\bf d}}
\newcommand{\ev}{{\bf e}}

\newcommand{\nv}{{\bf n}}

\newcommand{\pv}{{\bf p}}

\newcommand{\wv}{{\bf w}}

\newcommand{\xv}{{\bf x}}
\newcommand{\yv}{{\bf y}}

\newcommand{\Am}{{\bf A}}
\newcommand{\Bm}{{\bf B}}

\newcommand{\Dm}{{\bf D}}

\newcommand{\Id}{{\bf I}}

\newcommand{\Nm}{{\bf N}}

\newcommand{\Qm}{{\bf Q}}

\newcommand{\Wm}{{\bf W}}

\newcommand{\Ym}{{\bf Y}}

\newcommand{\gammav}{\hbox{\boldmath$\gamma$}}

\newcommand{\etav}{\hbox{\boldmath$\eta$}}

\newcommand{\epsilonv}{\hbox{\boldmath$\epsilon$}}

\newcommand{\muv}{\hbox{\boldmath$\mu$}}

\newcommand{\thetav}{\hbox{$\boldsymbol\theta$}}

\newcommand{\varthetav}{\hbox{\boldmath$\vartheta$}}
\newcommand{\varphiv}{\hbox{\boldmath$\varphi$}}

\newcommand{\Gammam}{\hbox{\boldmath$\Gamma$}}

\newcommand{\Phim}{\hbox{\boldmath$\Phi$}}

\newcommand{\Thetam}{\hbox{\boldmath$\Theta$}}

\newcommand{\trace}{{\hbox{tr}}}

\renewcommand{\arg}{{\hbox{arg}}}

\newcommand{\herm}{{\sf H}}

\begin{document}

\title{Mismatch Analysis and Cooperative Calibration of Array Beam Patterns for ISAC Systems}





\author{
Hui~Chen,~\IEEEmembership{Member,~IEEE},
Mengting~Li,~\IEEEmembership{Member,~IEEE},
Alireza~Pourafzal,~\IEEEmembership{Member,~IEEE},
Huiping Huang,~\IEEEmembership{Member,~IEEE},
Yu~Ge,~\IEEEmembership{Member,~IEEE},
Sigurd Sandor Petersen,
Ming Shen,~\IEEEmembership{Senior Member,~IEEE},
\\George C. Alexandropoulos,~\IEEEmembership{Senior Member,~IEEE},
and~Henk~Wymeersch,~\IEEEmembership{Fellow,~IEEE}

\thanks{H.~Chen, M.~Li, A.~Pourafzal, H.~Huang, Y. Ge and H.~Wymeersch are with the Department of Electrical Engineering, Chalmers University of Technology, 412 58 Gothenburg, Sweden (Email: {hui.chen; limeng; alireza.pourafzal; huiping; yuge; henkw}@chalmers.se).}
\thanks{M.~Li, S. S. Petersen and M. Shen are with the Aalborg University, Denmark (Email: {mengli; mish}@es.aau.dk, sp19@student.aau.dk).}
\thanks{G. C. Alexandropoulos is with the Department of Informatics and Telecommunications, National and Kapodistrian University of Athens, 16122 Athens, Greece (Email: alexandg@di.uoa.gr).}
\thanks{This work was supported, in part by the research grant (VIL59841) from VILLUM FONDEN, the Swedish Research Council (VR grant 2022-03007), the SNS JU project 6G-DISAC under the EU’s Horizon Europe research and innovation programme under Grant Agreement No 101139130, Vinnova B5GPOS Project under Grant 2022-01640, and the Chalmers Area-of-Advance Transport (Project ID: 95418010).}
}



\maketitle

\begin{abstract}
Integrated sensing and communication (ISAC) is a key technology for enabling a wide range of applications in future wireless systems. However, the sensing performance is often degraded by model mismatches caused by geometric errors (e.g., position and orientation) and hardware impairments (e.g., mutual coupling and amplifier non-linearity). This paper focuses on the angle estimation performance with antenna arrays and tackles the critical challenge of array beam pattern calibration for ISAC systems. To assess calibration quality from a sensing perspective, a novel performance metric that accounts for angle estimation error, rather than beam pattern similarity, is proposed and incorporated into a differentiable loss function. Additionally, a cooperative calibration framework is introduced, allowing multiple user equipments to iteratively optimize the beam pattern based on the proposed loss functions and local data, and collaboratively update global calibration parameters. The proposed models and algorithms are validated using real-world beam pattern measurements collected in an anechoic chamber. 
Experimental results show that the angle estimation error can be reduced from {$\textbf{1.01}^\circ$} to $\textbf{0.11}^\circ$ in 2D calibration scenarios, and from $\textbf{5.19}^\circ$ to $\textbf{0.86}^\circ$ in 3D calibration ones.
\end{abstract}

\begin{IEEEkeywords}
ISAC, array beam pattern, cooperative calibration, mismatch analysis, stochastic gradient descent.
\end{IEEEkeywords}

\section{Introduction}

\begin{figure}[t]
\centering
\centerline{\includegraphics[width=1.\linewidth]{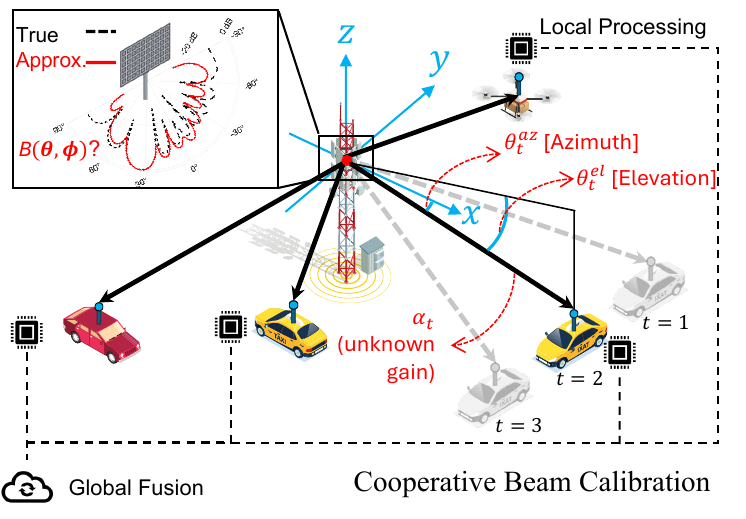}}
\vspace{-0.1cm}
\caption{Illustration of the considered cooperative beam calibration scenario. Local calibrated array beam patterns are obtained from cooperative UEs and then fused globally.}
\label{fig_illustration}
\vspace{-0.5cm}
\end{figure}

\Ac{isac} is becoming a central pillar of next-generation wireless systems to provide location information and environment awareness~\cite{liu2022integrated}.
With large array apertures and wide bandwidths, the communication signals enable not only higher data rates but also finer angular and delay resolutions~\cite{chen2022tutorial}. 
In addition to traditional \ac{toa} and \ac{tdoa} methods, angle-based measurements, such as \ac{aoa} and \ac{aod}, are now integrated into standardized systems like 3GPP NR~\cite{3GPP}. Such angle measurements provide substantial improvements in localization coverage and accuracy and enable complex tasks, including multipath-aided single \ac{bs} localization~\cite{mendrzik2018harnessing}, device orientation estimation~\cite{zheng2023coverage}, and \ac{slam}~\cite{lotti2023radio}.
These developments point toward a future in which 6G networks will support a diverse array of ISAC-driven services, from extended reality to autonomous navigation~\cite{stylianopoulos2025distributed}.

Despite the significant potential of ISAC, most current studies assume ideal system models and overlook the effects of calibration errors, such as geometric error~\cite{zheng2023jrcup} and hardware impairments~\cite{chen2023modeling}. While these assumptions facilitate performance analysis and the development of low-complexity algorithms (including super-resolution techniques for channel parameter estimation~\cite{wen2018tensor, pourafzal2025flex}), model mismatches lead to system performance degradation. Inaccurate beamforming caused by such mismatches reduces array gain, increases interference, and degrades link reliability~\cite{Sankar2024}. In sensing applications, the consequences are even more pronounced, as angular errors amplify with distance and can severely degrade system performance.\footnote{The \ac{snr} degradation caused by an angle-estimation error $\Delta\theta$ in an $N \times 1$ \ac{ula} with half-wavelength inter-element spacing is given by
$\mathrm{SNR}_{\text{loss}} = \left| \frac{1}{N} \sum_{n=0}^{N-1} e^{j \pi n \left( \sin\theta - \sin(\theta + \Delta\theta) \right)} \right|^2$, while the corresponding location error due to this angular error at distance $r$ is
$\Delta x = r \left| \tan(\theta + \Delta\theta) - \tan(\theta) \right|$. For a user at $\theta=0^\circ$ and $r=100~\mathrm{m}$ with $\Delta\theta=1^\circ$ and $N=8$, the SNR degradation and distance error are $0.069~\mathrm{dB}$ and $1.75~\mathrm{m}$, respectively.}
These inaccuracies affect the performance of channel parameter extraction and subsequent high-level tasks such as localization and mapping~\cite{ghazalian2025calibration}. This indicates that array calibration is needed to fully harness the potential of communication systems for sensing.

\begin{table*}[t]
    \caption{Summary of Beam Pattern Calibration Methods}
    {
    \centering
    \footnotesize 
    \resizebox{18cm}{!}{
    \begin{tabular}{c|c|c|c|c}
    \hline\hline
    Ref. & Purpose (Calibration Focus) & Hardware Type (Array Size) & Scenario & Key Methodology \\
    \hline    \hline
    \cite{pan2023situ} & 5G array error calibration & Digital array ($4$-element) & In situ & Expectation–maximization with interference cancellation 
    \\
    \cite{pohlmann2022simultaneous} & SLAC: joint position + calibration & Digital array (up to 4) & Self-calibration & Bayesian filtering, orientation, bias estimation \\
    \cite{takahashi2006theoretical} & Beam pattern characterization & Phased array ($4\times5$) & Offline & REV method (element-wise phase sweep) 
    \\
    \cite{takahashi2008fast} & Fast REV-based calibration & Phased array ($4\times 5$) & Offline & Enhanced REV (multi-element switching) \\
    \cite{takahashi2012novel} & Amplitude-only calibration & Phased array ($4\times 5$) & Offline & Fourier-based recovery from amplitude sweeps \\
    \cite{long2017fast} & Amplitude-only fast calibration & Phased array ($8$-element) & Offline & 3-phase shift measurement (0°, 90°, 180°) 
    \\
    \cite{long2017multi} & Simultaneous element calibration & Phased array ($8$-element) & Offline & Linear system solver using all-on measurements 
    \\
    \cite{zhang2019improved} & Reduced measurement load & Phased array ($4\times8$) & Offline & Sparse measurement, linear recovery 
    \\
    \cite{wang2021over} & OTA beam-mode calibration & Phased array ($4\times4$) & Offline & Calibration in beam-steering mode only 
    \\
    \cite{gradoni2021end} & Mutual coupling-aware modeling
    & RIS (up to $4\times 8$) 
    & Self-calibration & Electromagnetic end-to-end circuit model 
    \\
    \cite{zhang2023over} & OTA phase calibration & RIS ($8\times8$) 
    & In situ
    & Alternating block descent, CRB benchmark 
    \\
    \cite{fadakar2025mutual} & MC-aware localization & RIS ($20\times20$) & Self-calibration & Alternating optimization for joint localization and calibration 
    \\
    \cite{zhou2022transfer} & Excitation vector learning & Phased array (up to $4\times 8$) 
    & Offline
    & Transfer-learning-based surrogate model (power-only) \\ 
    \cite{mateos2025model} & Antenna displacement calibration & Phased array ($8\times 8$) & Self-calibration & Model-driven learning-based ISAC under hardware impairment \\
    \hline
    Ours & Sensing-based calibration & Phased array (up to $16\times16$) & In situ & Response and angle error loss based optimization\\
    \hline
    \end{tabular}}\par}
\label{tab:summary}
\end{table*}

\subsection{Related Works}
Array calibration scenarios can be broadly categorized into offline calibration (e.g., in chamber or controlled environment with dedicated calibration process), in situ calibration with known calibration agent state, and self-calibration that can calibrate the array without additional devices~\cite{pan2023situ}, as shown in Table~\ref{tab:summary}. Generally, offline calibration achieves the highest accuracy but requires specialized setups, whereas in situ and self-calibration methods are more practical in real deployments, with the former relying on a dedicated calibration agent.
In terms of calibration methods, they can be classified into element pattern calibration, amplitude-only calibration, complex-valued/structure-aware calibration, and \ac{ai}-based calibration. In the following, we review the state-of-the-art in each category and discuss their applicability and limitations for ISAC systems.

\subsubsection{Element Pattern Calibration}
Element calibration focuses on estimating element-wise antenna response. In~\cite{pan2023situ}, the calibration problem is framed as the estimation of an array error function (gain and phase error for each antenna), with experimental validation on a $4\times 1$ antenna array. 
With the cooperation of multiple mobile agents,~\cite{pohlmann2022simultaneous} presents a simultaneous localization and calibration (SLAC) approach that uses Bayesian filtering to estimate positions and calibration parameters in mobile networks jointly. 
While effective, such methods are generally limited to small-scale or linear arrays and do not scale efficiently to large or hybrid analog-digital arrays.


\subsubsection{Amplitude-Only Calibration}
Amplitude-only calibration methods, such as the widely-used \ac{rev} method and its variations~\cite{takahashi2006theoretical, takahashi2008fast, takahashi2012novel}, enable antenna array calibration by introducing additional phase shifters to perform element-wise phase sweeps and reconstruct the array response using only amplitude measurements. Enhanced REV methods, such as multi-element switching~\cite{takahashi2008fast} and Fourier-based recovery~\cite{takahashi2012novel}, further reduce calibration time and complexity. The fast amplitude-only calibration in~\cite{long2017fast} employs three-phase shift measurements (0°, 90°, 180°). Amplitude-only methods are particularly valuable when accurate phase measurements are unavailable or impractical; however, they fail to fully restore the complex beam pattern, limiting their effectiveness for angle estimation and high-precision sensing. Additionally, these methods often require hardware modifications (such as extra phase shifters), which may not be feasible for all array architectures, especially in the context of hybrid or distributed ISAC deployments.

\subsubsection{Complex-Valued and Structure-Aware Calibration}  
The third category of calibration techniques leverages linear algebraic structures and advanced modeling to reduce the number of required calibration measurements or to account for mutual coupling and hardware impairments. For example,~\cite{long2017multi, zhang2019improved, wang2021over} exploit linear system formulations and sparse recovery methods for calibration, reducing measurement overhead. In the context of \acp{ris}, impedance-based modeling has been proposed to describe mutual coupling~\cite{gradoni2021end}, while alternating optimization-based approaches have demonstrated high performance in~\ac{ota} test with complex propagation environments~\cite{zhang2023over} and joint localization and calibration scenarios~\cite{fadakar2025mutual}. These methods can provide accurate calibration in the presence of hardware non-idealities, but they often require access to detailed hardware parameters, precise control over system operation, or additional measurement campaigns. Furthermore, the assumptions made regarding array geometry, mutual coupling, or propagation environment may not generalize well to large-scale, distributed, or hybrid analog-digital systems.

\subsubsection{AI-Based Calibration}  
Recent advances in data-driven calibration approaches aim to address practical challenges in dynamic and distributed environments. In~\cite{zhou2022transfer}, a surrogate-model-based method is introduced for phased array calibration using only power measurements, leveraging transfer learning to adapt excitation vector models across devices. Model-driven deep learning frameworks have also shown promise for correcting antenna displacement errors~\cite{mateos2025model}. These AI-based methods are attractive for large-scale and practical deployments, as they can learn from real-world data and adapt to changing conditions. However, they typically focus on beam pattern reconstruction accuracy and do not directly address the impact of calibration on sensing accuracy. Moreover, these approaches often require extensive training data and may not be robust to all types of hardware impairments or environmental dynamics.

While the above calibration methods offer valuable solutions for specific scenarios, most existing works evaluate calibration quality based on beam pattern reconstruction accuracy, without explicitly quantifying the impact of beam mismatch on sensing or localization performance. Additionally, much of the literature assumes fully digital architectures or fixed array types, overlooking the analog and hybrid beamforming structures and distributed calibration needs common in practical \ac{mmwave} ISAC systems. These limitations reduce the practical relevance of current calibration methods for real-world sensing and localization tasks.

\subsection{Contributions}
In this work, we address in situ beam pattern calibration using downlink pilot signals from a sensing-oriented perspective: we quantify how beam mismatch degrades localization accuracy, and we develop both a practical beam pattern model and a scalable cooperative calibration algorithm suitable for distributed implementation under hybrid beamforming constraints.

The main contributions of this paper are as follows:
\begin{itemize}
    \item We formulate a cooperative beam calibration problem in a distributed \ac{isac} scenario. Array calibration of the \ac{bs} is performed at each \ac{ue} via communication pilot signals, and multiple \acp{ue} collaborate to jointly calibrate the beam pattern of the \ac{bs} antenna array, as illustrated in Fig.~\ref{fig_illustration}.
    \item We develop a novel, sensing-oriented performance metric, obtained by calculating the pseudo-true parameter that minimizes the Kullback-Leibler divergence between the actual model and the mismatched model after calibration. Unlike traditional beam pattern similarity metrics, the proposed metric better reflects calibration quality in terms of sensing performance.
    \item We propose an iterative optimization algorithm adopting a differentiable objective function, approximated from the sensing-based performance metric. The algorithm can further refine communication-based calibration for better sensing performance and support cooperation between multiple calibration agents.
    \item We validate the proposed beam pattern models and calibration algorithms based on real-world beam pattern measurements in an anechoic chamber. Both communication- and sensing-oriented performance metrics are assessed across various scenarios and hyperparameter settings.
\end{itemize}

\textit{Notations and Symbols:} Italic letters denote scalars (e.g., $a$), bold lower-case letters denote vectors (e.g., $\av$), and bold upper-case letters denote matrices (e.g., $\Am$). $(\cdot)^\top$, $(\cdot)^\herm$, $(\cdot)^*$, $(\cdot)^{-1}$, $\trace(\cdot)$, $\Vert{\cdot}\Vert$, and $\Vert{\cdot}\Vert_\text{F}$ represent the transpose, Hermitian transpose, complex conjugate, inverse, trace, $\ell$-2 norm, and Frobenius norm operations, respectively; $\Am \odot \Bm$, $\Am \oslash \Bm$, $\Am \otimes \Bm$, $\av \circ \bv$ are the Hadamard product, Hadamard division, Kronecker product, and outer product, respectively; $[\cdot,\ \cdot,\ \cdots, \cdot]^\top$ denotes a column vector; $[\cdot]_{i,j}$ is the element in the $i$-th row, $j$-th column of a matrix, and $[\cdot]_{a:b,c:d}$ is the submatrix constructed from the $a$-th to the $b$-th row, and the $c$-th to $d$-th column of a matrix; $\mathrm{Re}\{a\}$ and $\mathrm{Im}\{a\}$  extracts the real and imaginary parts of a complex variable, $\angle(a)$ denotes the phase of a complex scalar $a$; $\bm{1}_{N}$ denotes an $N\times 1$ all-ones vector, and $\mathbf{I}_{N}$ denotes a size-$N$ identity matrix. 

\section{System Model}

In this section, we begin with a general system model that includes geometrical information about the transmitter and receiver, as well as the beam pattern. Then, several beam representations will be described, followed by the corresponding calibration models.

\subsection{Generic System Model}
We consider a far-field \ac{ofdm}-based downlink system with one \ac{bs} and multiple \acp{ue}. For simplicity, we describe the system model for a specific \ac{bs}-\ac{ue} link, and thus the UE index is dropped throughout this section. The BS located at $\pv_\text{B}\in \mathbb{R}^3$ is equipped with an analog uniform planar array consisting of $N=N_r \times N_c$ antennas ($N_r$ rows and $N_c$ columns), and a single-antenna UE moves following a predefined trajectory (e.g., from $\pv_{\text{U},1}\in \mathbb{R}^3$ to $\pv_{\text{U},T}$ with a total $T$ snapshots) and is receiving pilot signals from the \ac{bs}.\footnote{This work uses downlink signals to highlight cooperative calibration in a distributed system, and the feedback between UEs and BS is needed to provide calibrated parameters. The calibration problem can also be performed using uplink signals with proper interference management.} 
At snapshot $t$, $G$ \ac{ofdm} symbols are transmitted from \ac{bs} to \ac{ue}, each with $K$ subcarriers and a unique beamforming vector (forming a $G$ codewords beam sweeping), and the received signals $\Ym_t\in \mathbb{C}^{G\times K}$ can be expressed as the combination of signals propagating $L+1$ paths given by
\begin{equation}
    \Ym_t = \sum_{\ell=0}^L \alpha_{\ell, t} \bv(\varthetav_{\ell, t}) (\dv(\tau_{\ell, t})
    \odot \xv_t)^\top + \Nm_t,
    \label{eq_mismatched_system_model}
\end{equation}
where $\ell$ is the index of path with $\ell=0$ denoting the \ac{los} path, $\alpha_{0, t} = \sqrt{P}\frac{\lambda}{4\pi r_{0, t}}e^{j \beta_{0,t}}$ is the LOS complex channel gain with $P$ as the transmit power, $\beta_{0, t}$ as the phase, and $r_0=\Vert \pv_\text{B} - \pv_\text{U}\Vert$ as the distance between the $t$-th measurement location and the BS array. 
The channel gain of the \ac{nlos} path ($\ell>0$) can be expressed as $\alpha_{\ell, t} = \sqrt{P}\frac{c_\ell}{\sqrt{4\pi}}\frac{\lambda}{4\pi r_{\ell, t, 1}r_{\ell, t, 2}}e^{j \beta_{\ell,t}}$, with $c_\ell$ as the radar cross-section coefficient, $r_{\ell, t, 1} = \Vert \pv_{\text{B}} - \pv_{\text{S}, \ell, t} \Vert$ and $r_{\ell, t, 1} = \Vert \pv_\text{U} - \pv_{\text{S}, \ell, t} \Vert$ denoting the distances from the BS and UE to the scattering point (located at $\pv_{\text{S}, \ell, t}$), respectively.
Angle pair $\varthetav_{\ell, t} = [\theta^{\mathrm{az}}_{\ell, t}, \theta^{\mathrm{el}}_{\ell, t}]^\top $ is the \ac{aod} in both azimuth and elevation, $\tau_{\ell, t}$ is the signal delay. The pilot signal vector is denoted by $\xv_{t}\in \mathbb{C}^{K}$ with unit norm as $|x_{t,g}| = 1$, and $\Nm_t\in \mathbb{C}^{G \times K}$ is the additive white Gaussian noise matrix with each element $n_{t,g}\in\mathcal{CN}(0, \sigma_n^2)$. The beam response vector (for $G$ codewords) at a specific angle pair $\varthetav$ is denoted as $\bv(\varthetav) = [b_{1}(\varthetav), \ldots, b_{G}(\varthetav)]^\top \in \mathbb{C}^{G}$, which will be detailed in the following subsection, and $\dv(\tau) \in \CC^{K}$ reflects the phase offset across different subcarriers with each element $d_k(\tau) = e^{j2\pi k \Delta_f \tau}$, where $\tau$ is the delay and $\Delta_f$ is the subcarrier spacing.
A detailed geometric relationship between the BS/UE position, orientation, and channel parameters can be found in~\cite{chen2022tutorial}.



{By de-rotating the phase caused by the LOS delay $\hat\tau_{0,t}$, removing the pilot signals as $\Ym_t\oslash (\bm{1}_{_G} \xv_t ^\top)$, and performing coherent combining the $K$ subcarriers, the generic system model can be simplified from~\eqref{eq_mismatched_system_model} as
\begin{equation}
    \yv_t = \underbrace{\tilde\gamma_t \bv(\varthetav_t)}_{\text{LOS path}}+ \underbrace{\sum_{\ell=1}^L \alpha_{\ell, t} \bv(\varthetav_{\ell, t}) \dv^\herm(\hat\tau_{0, t}) \dv(\tau_{\ell, t})}_{\text{NLOS paths}} + \nv_t.
    \label{eq_compact_vector_form}
\end{equation}
Here, $\tilde \gamma_t = \tilde \gamma_{0,t} = \alpha_{\ell, t}\dv^\herm(\hat\tau_{0,t})\dv(\tau_{0,t})$ with $\hat \tau_{0,t}$ as the estimated LOS delay, and the noise level is scaled as $n_{t,g}\in \mathcal{CN}(0, K\sigma_n^2)$.}
We further assume that the bandwidth is sufficiently wide to resolve the delay of the LOS path and NLOS paths~\cite{pourafzal2025flex}, and the NLOS path term in~\eqref{eq_compact_vector_form} approximates to zero.
By concatenating beam pattern vectors for all the codewords into a matrix $\Bm(\Thetam) \in \mathbb{C}^{G\times T}$ with $\Thetam = [\varthetav_1, \ldots, \varthetav_T]$, we obtain 
\begin{equation}
    \Ym = \Bm(\Thetam) \tilde\Gammam + \Nm.
    \label{eq_compact_matrix_form}
\end{equation}
Here, $\Ym \in \mathbb{C}^{G\times T}$ contains all the measurements from $T$ position/angles for a specific UE, and $\tilde\Gammam = \text{diag}{([\tilde\gamma_1, \ldots, \tilde\gamma_T])}$ contains the LOS complex channel gain for each measurement.

\subsection{Beam Representation Model}
The goal of beam representation is to find a simplified model to represent $\Bm(\Thetam)$ and characterize the ground truth beam patterns (e.g., measured from a chamber). We adopt two beam representation models, namely, an ideal model (steering-based codebook with fixed beamforming directions) and a practical model (learned codebook). The former model is widely used in existing works~\cite{liu2022integrated, zheng2023coverage, zheng2023jrcup, chen2023modeling, wen2018tensor} and will be used as a benchmark for the latter practical model.

\subsubsection{Ideal Beam Model}
Without loss of generality, we define an ideal array response at angle $\varthetav$ as
\begin{equation}
    \tilde\bv{(\varthetav)} = g(\varthetav) {\bar \Wm^\herm(\bar\Phim)} \av(\varthetav),
    \label{eq_ideal_beam_response_vector}
\end{equation}
where $g(\varthetav)$ is the element pattern.\footnote{At the UE side, element pattern should also be considered. However, we will show in Sec.~{\ref{Sec_sensing_model}} that the element patterns of both UE and BS do not affect sensing performance and hence are treated as nuisance parameters.} 
A simple patch antenna model can be implemented as~\cite{balanis2016antenna} 
\begin{equation}
g(\varthetav, \beta) = \cos^{\beta}(\theta_\mathrm{eff}) = (\cos\theta^\mathrm{az}\cos\theta^{\mathrm{el}})^\beta
\label{eq_element_pattern}
\end{equation}
with $\theta_\mathrm{eff} = \cos^{-1}(\cos\theta^\mathrm{az}\cos\theta^{\mathrm{el}})$ being the effective angle between the target direction and the array boresight, and $\beta>0$ is the directivity parameter. However, the following calibration tasks are not limited to the beam pattern model defined in~\eqref{eq_element_pattern}. The steering vector $\mathbf{a}(\varthetav)$ captures the phase differences due to array configuration across different antennas (assuming half-wavelength spacing) as
\begin{align}
    \mathbf{a}(\varthetav) = \mathbf{a}_c(\theta^{\mathrm{az}}, \theta^{\mathrm{el}}) \otimes \mathbf{a}_r(\theta^{\mathrm{el}}),
\end{align}
with
\begin{align}
\mathbf{a}_c(\theta^{\mathrm{az}}, \theta^{\mathrm{el}}) & = 
[1,\ \cdots,\  e^{j  \pi (N_c-1) \sin\theta^{\mathrm{az}} \cos\theta^{\mathrm{el}}}]^\top,
\\
\mathbf{a}_r(\theta^{\mathrm{el}}) & = 
[1,\  \cdots ,\  e^{j  \pi (N_r-1) \sin\theta^{\mathrm{el}}}]^\top.
\end{align}
In the ideal model, each codeword in the precoding matrix (or codebook) $\bar \Wm(\Phim) = [\wv(\bar\varphiv_1), \ldots, \wv(\bar\varphiv_G)] \in \mathbb{C}^{N\times G}$ is defined the same as the steering vector (e.g., $\wv(\varphiv) = \av(\varphiv)$) with the constraints $|w_{g, n}| = 1$, and $\varphiv = [\phi^\mathrm{az}, \phi^\mathrm{el}]^\top$ is the beamforming direction. Thus, each codeword maximizes the \ac{snr} at a certain direction $\bar\varphiv_g$, and $\bar\Wm$ can be represented by a set of beamforming angles $\bar\Phim = [\bar\varphiv_1, \ldots, \bar\varphiv_G] \in \mathbb{C^{2\times G}}$ (predefined by the array manufacturer).

\subsubsection{Practical Beam Model}
As investigated in our initial work in \cite{mengli2025ris}, an ideal codebook can lead to severe mismatch and performance degradation. Alternatively, a practical codebook $\Wm$ (with each codeword satisfying $\Vert \wv_g\Vert = 1$) containing hardware limitations and realization issues can be defined as\footnote{Note that the definition of $\Wm$ could be either $\bar\Wm(\bar\Phim) + \Delta_\Wm$ or $\bar\Wm(\bar\Phim) \odot \Delta_\Wm$. Since we estimate $\Wm$ as a whole without access to $\Delta_{\Wm}$, the choice of these two forms does not affect calibration.}
\begin{equation}
    \Wm = \bar\Wm(\bar\Phim) + \Delta_\Wm.
\end{equation}
Hence, the beam pattern model of a specific direction can further be defined as 
\begin{equation}
    \bv{(\varthetav)} = g(\varthetav){\Wm}^\herm \av(\varthetav).
    \label{eq_model_representation}
\end{equation}
While more complex impairment models exist (e.g., array gain error considered in~\cite{chen2023modeling}), the formulation in~\eqref{eq_model_representation} is an approximation that yields accurate calibration performance with reduced complexity, which will be the main focus of this work. 

\subsection{Sensing Model}
\label{Sec_sensing_model}
Given a beam representation model, the sensing task can be defined as extracting the angle $\varthetav$ from the observed signal $\yv$. Considering the beam pattern error has a limited effect on delay estimation~\cite{chen2023modeling}, we focus on angle estimation based on \ac{mle}. We show that based on the model in~\eqref{eq_compact_matrix_form}, sensing relies only on the codebook $\Wm$, and the channel gain matrix $\Gammam$ is a nuisance parameter for sensing-based calibration.

The angle estimation for the LOS scenario can be formulated as~\cite{chen2024multi}
\begin{equation}
    [\varthetav, \alpha] = \argmin_{\boldsymbol{\vartheta}, \alpha} \Vert \yv - \alpha\bv(\varthetav)\Vert,
    \label{eq_maximum_likelihood}
\end{equation}
where $\yv$ is an observation vector for a specific location. To simplify the estimation, the nuisance parameter $\alpha$ can be represented as $\alpha = \frac{\bv^\herm(\boldsymbol{\vartheta})\yv}{\bv^\herm(\boldsymbol{\vartheta})\bv(\boldsymbol{\vartheta})}$ based on~\eqref{eq_maximum_likelihood} with a given $\varthetav$, and the estimation problem can be written as~\cite{chen2024multi}
\begin{equation}
\begin{split}
\varthetav 
& = \argmin_{\boldsymbol{\vartheta}} \| \yv - \frac{\bv^\herm(\boldsymbol{\vartheta})\yv}{\bv^\herm(\boldsymbol{\vartheta})\bv(\boldsymbol{\vartheta})}\bv(\varthetav) \|^2
\\
& = \argmin_{\boldsymbol{\vartheta}} \yv^\herm\yv - \frac{\bv^\herm(\varthetav)\yv}{\bv^\herm(\varthetav)\bv(\varthetav)}\yv^H\bv(\varthetav)
\\
& = \argmax_{\boldsymbol{\vartheta}} \frac{|\bv^\herm(\varthetav) \yv|}{\Vert\bv(\varthetav)\Vert} = \argmax_{\boldsymbol{\vartheta}} \frac{|\av^\herm(\varthetav)\Wm \yv|}{\Vert \Wm^\herm \av(\varthetav)\Vert}.
\end{split}
\label{eq_joint_theta_gamma}
\end{equation}
From~\eqref{eq_joint_theta_gamma} we can see that angle estimation is not related to the element pattern as $g(\varthetav)$. However, to reconstruct the beam pattern for communication purposes, estimation of the element pattern is needed.

\subsection{Calibration Models}
The goal of this calibration work is to find a beam representation $\bv(\varthetav)$ defined in~\eqref{eq_model_representation}, based on the received pilot signals $\Ym$ in~\eqref{eq_compact_matrix_form}, such that the sensing performance in~\eqref{eq_maximum_likelihood} can be improved. Next, we first describe the compact calibration model, followed by several calibration models. 
To assist calibration, we incorporate the element pattern (originally from $\bv(\varthetav_t)$ as shown in~\eqref{eq_model_representation}) into the gain matrix $\tilde \Gammam$ to form $\Gammam = \text{diag}(\gammav)$ with each diagonal element as $\gamma_t = \tilde\gamma_t g(\varthetav_t)$. Consequently, the generic system model can be reformulated from~\eqref{eq_compact_matrix_form} as 
\begin{equation}
    \Ym = \Wm^\herm \Am(\Thetam) \Gammam + \Nm,
    \label{eq_compact_matrix_form_2}
\end{equation}
with $\Am(\Thetam) = [\av(\varthetav_1), \ldots, \av(\varthetav_T)] \in \mathbb{C}^{N\times T}$ as the steering matrix.
Once the codebook $\hat\Wm$ and the gain matrix $\hat\Gammam$ are obtained, the calibrated beam response can be expressed as 
\begin{equation}
    \hat\bv(\varthetav) = \hat g(\varthetav)\hat\Wm^\herm \av(\varthetav).
\end{equation}
Since the gain estimate $\Gammam$ corresponds to certain angles, we interpolate 
the element pattern $\hat g(\varthetav_t) = r_t |\gamma_t|$ to obtain $\hat g(\varthetav)$, where $r_t$ is the compensation of path loss. Note that the beam response $\bv(\varthetav)$ reflects the complex response gain for different angles, and normalization will be performed when evaluating the calibration accuracy.




In this work, we assume UEs are dedicated calibration agents with known location~\footnote{If UE positions are unknown, calibration (i.e., estimating $\Wm$ and $\Gammam$) needs to be performed jointly with localization (i.e., estimating $\Thetam$). However, this joint problem is non-identifiable, as the observation model is invariant to unitary transformations of the latent factors. For any for any unitary matrix $\Dm$, $(\Dm\Wm)^\herm\Dm\Am = \Wm^\herm\Am$, yielding equivalent observations.
With appropriate priors (e.g., a partially calibrated codebook $\Wm$ or known reference directions), the joint estimation can be reformulated as a Bayesian inference problem, which is left for future work.}
in each of the $T$ measurements (e.g., from GPS), and hence the calibration task is to estimate the unknown parameters (e.g., codebook $\Wm$ and beam gain matrix $\Gammam$, depending on the scenarios) and reconstruct the beam pattern $\bv(\theta)$ for better communication and sensing performance.
Next, we detail different calibration scenarios, where model M1 works as the benchmark for models M2-M4, and the main focus is on M4 throughout this work.

\subsubsection{Benchmark Beam Pattern Model (M1)}
This model adopts an ideal codebook $\bar\Wm(\bar\Phim)$ with known beamforming directions $\bar\Phim$. We assume the sample angle $\varthetav_t$ and distance $r_t$ are known for each measurement point, the calibration objective is to estimate the channel gain matrix $\Gammam$ and then extract the element pattern as $g(\varthetav_t) = r_t |\gamma_t|$. The corresponding signal model can be expressed as
\begin{equation}
    \Ym^{\text{M1}} = {\bar\Wm^\herm(\bar\Phim) \Am(\bar\Thetam) \Gammam} + \Nm.
    \label{eq_model1}
\end{equation}
From sensing perspective, no calibration is needed for M1, and the $\Gammam$ is estimated for beam reconstruction purpose.

\subsubsection{Beamforming Angle Calibration (M2)}
Considering the inaccuracy of the beamforming angle provided by the array manufacturer, M2 focuses on the calibration of the matrices $\Phim$ and $\Gammam$ at the same time. The signal model in this case can be expressed as follows: 
\begin{equation}
    \Ym^{\text{M2}} = {\bar\Wm^\herm(\Phim) \Am(\bar\Thetam) \Gammam} + \Nm.
    \label{eq_model2}
\end{equation}
The calibration M2 is expected to provide better performance than M1 with more accurate beamforming angles.

\subsubsection{{Codebook Calibration (M3)}}
Codebook calibration can only be performed offline to remove the effect of channel gain (e.g., inside the chamber or a well-controlled environment).
Then, the calibration task becomes estimating the codebook $\Wm$ and the directionality coefficient $\beta$ defined in~\eqref{eq_element_pattern}, assuming UE locations are known, and the signal model can be expressed as \cite{zhang2024multiprobe,wang2021over}
\begin{equation}
    \Ym^{\text{M3}} = {\Wm^\herm \Am(\bar\Thetam) \Gammam(\bar\Thetam, \beta, \bar\dv)} + \Nm.
    \label{eq_model3}
\end{equation}
Note that M3 is the only model that uses the element pattern model defined in~\eqref{eq_element_pattern}. The purpose of this model is to highlight the coupling of channel gain and element pattern, and a simple element pattern cannot well-capture the beam response.



\subsubsection{Joint Codebook and Gain Matrix Calibration (M4)}
The most practical model is to estimate $\Wm$ and nuisance parameters $\Gammam$ jointly, and the calibration model can be formulated as~{\cite{wang2021over}}
\begin{equation}
    \Ym^{\text{M4}} = {\Wm^\herm \Am(\bar\Thetam) \Gammam} + \Nm.
    \label{eq_model4}
\end{equation}
Compared with~\eqref{eq_model1} to~\eqref{eq_model2}, the introduced unknowns $\Wm$ and $\Gammam$ increase the difficulties of the beam calibration task, requiring dedicated calibration algorithms, which will be detailed in Section~\ref{sec_calibration_algorithm}. 


\section{Performance Metrics and Loss Functions}
This section starts by introducing several performance metrics, namely, response similarity, bias angle error, and codebook gain loss. Then, response error loss and angle error loss will be explained for calibration.

\subsection{Performance Metrics}
To evaluate the quality of beam calibration, we introduce beam response similarity, angle estimation error, and codebook gain loss to evaluate the beam pattern error, sensing, and communication performances using the calibrated beams, respectively.


\subsubsection{Beam Response Similarity}
{The total variation distance of power angular spectrum is often used to quantify the similarity between the reconstructed beam pattern and the reference one in \ac{ota} and channel emulation testing, as defined in 3GPP TR 38.827~\cite{tr38827}. With the reconstructed pattern $\hat{\bv}(\varthetav)$ and the ground truth $\bar\bv(\varthetav)$, the variation distance of two beam patterns can be defined as
\begin{equation}
    \tilde{E}_\mathrm{R} = \frac{1}{2}\int_{\boldsymbol{\vartheta}} 
    \left|
    \frac{|\hat b(\varthetav)|}{\int_{\boldsymbol{\vartheta}}|\hat b(\varthetav)| \text{d}\varthetav} - 
    \frac{|\bar b(\varthetav)|}{\int_{\boldsymbol{\vartheta}}|\bar b(\varthetav)| \text{d}\varthetav}
    \right|
    \text{d}\varthetav.
    \label{eq_variation_distance}
\end{equation}}
However, the above equation only reflects the gain of the beam pattern (i.e., phase is ignored) and the integral is impractical for the discrete measurements. Based on~\eqref{eq_variation_distance}, we define a discrete version of the normalized response error $E_\mathrm{R}$. By choosing a set of evaluation angles $\Thetam_s = [\varthetav_1, \ldots, \varthetav_S]$ ($S\le T$ with $\Thetam_s$ as a subset of all the measurement angles $\Thetam$), the response similarity $S_\mathrm{R}$ ($0\le S_\mathrm{R}\le 1$) can be defined as 
\begin{equation}
    S_\mathrm{R} = 1 - E_\mathrm{R} = 1-\frac{1}{2}\left\Vert 
    \frac{\hat\Bm(\Thetam_s)}{\Vert \hat\Bm(\Thetam_s)\Vert_\text{F}} - \frac{\bar\Bm(\Thetam_s)}{\Vert \bar\Bm(\Thetam_s)\Vert_\text{F}}
    \right\Vert_\text{F}.
    \label{eq_loss_1}
\end{equation}
Here, the response similarity can also be modified by adding different weighting factors for each evaluation angle, depending on the area of interest (e.g., large weights for boresight directions). 
For simplicity, we set the weighting factor to be identical for all the measurement angles and yield~\eqref{eq_loss_1}.


\subsubsection{Angle Estimation Bias}
\label{sec:performance_metric_angle_error}
Although response similarity can be used to evaluate differences between reconstructed beam patterns and ground truth, its performance in angle estimation cannot be directly quantified.
To facilitate sensing-based performance evaluation, we adopt a pseudo-true parameter vector, which is defined as the point that minimizes the Kullback-Leibler divergence between $f_\text{TM}(\yv|\bar {\boldsymbol\eta})$ and $f_\text{MM}(\yv| {\boldsymbol\eta})$ as
\begin{align}
    {\boldsymbol\eta}_0 = \arg \min_{\boldsymbol\eta} D_\text{KL}(f_\text{TM}(\yv|\bar {\boldsymbol\eta})\Vert f_\text{MM}(\yv| {\boldsymbol\eta})).
\end{align}
with $f_\text{TM}(\yv|\bar {\boldsymbol\eta})$ and $f_\text{MM}(\yv| {\boldsymbol\eta})$ as the probability density functions of the true model and the mismatched model, respectively. The pseudo-true parameter is denoted as $\etav_0 = [\mathrm{Re}(\gamma_0), \mathrm{Re}(\gamma_0), \theta_0^{\text{az}}, \theta_0^{\text{el}}]$, containing the nuisance parameter channel gain and the angle estimation. The groundtruth state vector and the state variable are denoted as $\bar\etav$ and $\etav$.

Specifically, we adopt the ground truth beam pattern measurement $\bar\bv(\varthetav)$ from the chamber as the true model 
\begin{equation}
    \yv_\text{TM} = \underbrace{\gamma\bar\bv(\varthetav)}_{=\bar\muv(\etav)} + \nv, 
\end{equation}
and take the calibrated model as the mismatched model 
\begin{equation}
    \yv_\text{MM} = \underbrace{\gamma \bv(\varthetav)}_{=\muv(\etav)}  + \nv = \gamma \Wm^\herm \av(\varthetav)  + \nv,
    \label{eq_y_mm}
\end{equation} 
where $\yv_\text{TM} \sim f_{\text{TM}}( \yv|\etav)$ and $\yv_\text{MM} \sim f_{\text{MM}}( \yv|\etav)$. Note that the pseudo-true parameter reflects the best performance that an estimator can achieve, which is limited by the bias as $\Vert \varthetav_0 - \bar\varthetav\Vert$. 
By defining $\epsilonv(\etav) \triangleq \bar\muv(\bar {\boldsymbol\eta}) - \muv({\boldsymbol\eta})$, the pseudo-true parameter can be obtained as follows~\cite{ozturk2022ris}:
\begin{equation}
    {\boldsymbol\eta}_0 
    = \arg \min_{{\boldsymbol\eta}} \Vert \epsilonv(\etav) \Vert^2 
    = \arg \min_{{\boldsymbol\eta}} \Vert \bar\muv(\bar {\boldsymbol\eta}) - \muv({\boldsymbol\eta})\Vert^2  
    \label{eq_pseudotrue_final}.
\end{equation}
A practical solution is to estimate ${\boldsymbol\eta}_0$ using gradient-based methods initialized with the true value $\bar {\boldsymbol\eta}$ under the assumption that UE positions are known. 
With the pseudo-true angle estimation $\varthetav_0 = [\theta_0^{\text{az}}, \theta_0^{\text{ele}}]^\top$, we define angle estimation error as
\begin{equation}
    \tilde{E}_{\mathrm{A}} = \int_{\boldsymbol{\vartheta}} \Vert \hat\varthetav_{\yv_{\text{MM}|\hat\bv(\cdot)}}(\bar\bv(\varthetav)) - \varthetav \Vert^2 \text{d}\varthetav,
    \label{eq_angle_estimation_error_l2}
\end{equation}
where $\hat\varthetav_{\yv_\text{MM}|\hat\bv(\cdot)}(\bar\bv(\varthetav)) \triangleq \varthetav_0$ denotes the estimated angle with $\bar\bv(\varthetav)$ as the input and the reconstructed beam pattern $\hat\bv(\cdot)$ for signal model $\yv_{\text{MM}|\hat\bv(\cdot)}$. Similar to~\eqref{eq_loss_1}, a tractable angle error loss can be defined using a set of evaluation angles $\Thetam_s = [\varthetav_1, \ldots, \varthetav_S]^\top$ as
\begin{equation}
    {E}_{\mathrm{A}} = \frac{\sum_{s}  \Vert \hat\varthetav_{\yv_{\text{MM}|\hat\bv(\theta)}}(\bar\bv(\varthetav_s)) - \varthetav_s \Vert^2}{S}.
    \label{eq_loss2}
\end{equation}
Here, the angle error $E_{\mathrm{A}}$ captures the average angle estimation error using the calibrated beam pattern model to process the data from the true model, which serves as a sensing-oriented performance metric. 

\subsubsection{{Codebook Gain Loss}}
As a benchmark for communications, we define gain loss to quantify the effect of calibration on communication performance in the LOS scenario. Specifically, the best beam is selected based on the known beam pattern to achieve the maximum LOS gain (i.e., ideal beam pattern for the uncalibrated case and M1 to M4 for the calibrated array). Similar to the $E_\text{R}$ and $E_\text{A}$, the gain loss can be expressed as the average relative gain between the BS-assumed beam pattern and the ideal beam pattern among all the evaluation angles
\begin{equation}
    E_{\mathrm{C}} = \frac{1}{S}\sum_s\frac
    { 
    \left\| \bar{\bv}^\herm_{\hat{g}^{\star}_s}(\varthetav_s)\av(\varthetav_s)\right\|^2}
    {\left\| \tilde{\bv}^\herm_{\tilde{g}^{\star}_s}(\varthetav_s)\av(\varthetav_s)\right\|^2}.
    \label{eq_loss3}
\end{equation}
where $\hat{g}^{\star}_s\! =\! \argmax_g |{\hat\bv}_g(\varthetav_s)|$ and $\tilde{g}^{\star}_s\! =\! \argmax_g |{\tilde\bv}_g(\varthetav_s)|$ are the optimal beam indices obtained from the calibrated beam pattern $\hat\bv(\varthetav)$ and the ideal beam pattern $\tilde\bv(\varthetav)$ defined in~\eqref{eq_ideal_beam_response_vector}, respectively. 
The above equation captures the weighted gain loss compared with an ideal array beam pattern, which is used for evaluating communication performance improvement after calibration.

\subsection{Loss Functions}
\subsubsection{Response Error Loss}
Inspired by the response similarity defined in~\eqref{eq_loss_1}, we aim to reduce the error between the reconstructed beam pattern and the reference one, and formulate the \ac{rel}-based calibration. Given the received signal $\Ym$, the calibration task is to find the optimal $\Wm$ and $\Gammam$ that minimize the \ac{rel} $L_\text{REL}$ as
\begin{equation}
    [\hat{\Wm}, \hat{\Gammam}] 
    = \arg\min_{\Wm,\,\Gammam} \; L_{\mathrm{R}}(\Ym, \Wm, \Gammam),
    \quad \text{s.t.} \|\wv_g\|_2 = 1, \; \forall g,
\end{equation}
where 
\begin{equation}
L_\mathrm{R}(\Ym, \Wm, \Gammam)=\| \Ym - \Wm^\herm \Am \boldsymbol{\Gamma} \|_\text{F}^2/T.
\label{eq_REL_loss_function}
\end{equation}
The above formulation can be implemented for both 2D and 3D beams, with each column of $\Am$ representing a steering vector for a specific target position.


\subsubsection{Angle Error Loss}
Based on the definition of angle error in~\eqref{eq_loss2}, we can formulate \ac{ael} as
\begin{equation}
    \tilde{L}_{\mathrm{A}}(\Ym, \Wm, \Gammam) = {\sum_{t=1}^T \Vert \hat\varthetav_{\yv_{\text{MM}|\hat\bv(\theta)}}(\yv_t) - \varthetav_t \Vert^2}/T.
    \label{eq_AEL_loss_function}
\end{equation}
However, this loss function requires bilevel optimization~\cite{zhang2024introduction}, where an inner optimization (i.e., finding the pseudo-true angle) needs to be solved to calculate the loss values. To assist calibration, we propose a differentiable \ac{ael} function to reduce sensing angle error.


From~\eqref{eq_joint_theta_gamma}, it is shown that the angle estimation is trying to find the maximized normalized projection of the beam response onto the observations. Inspired by this, we define a new loss function as an alternative to $L_2$ by matching the normalized projection of the observations as
\begin{equation}
\begin{split}
    & L_{\mathrm{A}}(\Ym, \Wm, \Gammam) = 
    \frac{\sum_{t,s} |e_{t,s}|^2}
    {TS},
    \\
    & e_{t,s}  = u_{t,s} - \check{u}_{t,s} = 
    \frac{\bv^\herm(\varthetav_t)\yv_s}{\Vert \bv(\varthetav_t)\Vert} 
    - \frac{\yv_t^\herm\yv_s}{\Vert \yv_t\Vert}.
\end{split}
\label{eq_loss_function_3}
\end{equation}
Here, $\bv(\varthetav) = \gamma(\varthetav) \Wm^\herm \av(\varthetav)$ denotes the calibrated beam pattern, and $\yv_t$ and $\yv_s$ are the received signal at the $\varthetav_t$ and $\varthetav_s$, respectively. For the ease of gradient calculation, we adopt ${\bv^\herm(\varthetav_t)\yv_s}$ instead of ${|{\bv^\herm(\varthetav_t)\yv_s}|}$. Similar to~\eqref{eq_loss_1}, the evaluation angles $\Thetam_s = [\varthetav_1, \ldots, \varthetav_S]^\top$ are chosen as a subset of the measurement angles $\Thetam = [\varthetav_1, \ldots, \varthetav_T]^\top$ to reduce the complexity of the algorithm. 
The term $\check u_{t,s}$ is defined by replacing $\bar{\bv}(\varthetav_t)$ in $\frac{\bar\bv^\herm(\varthetav_t)\yv_s}{\Vert \bar\bv(\varthetav_t)\Vert}$ with a noisy measurement $\yv_t$ for practical implementation purposes.
With the loss function defined in~\eqref{eq_loss_function_3}, the problem can be solved using iterative optimization.

\section{Calibration Algorithm}
\label{sec_calibration_algorithm}
In this section, we start with the \ac{rel}-based calibration algorithm for model M4 using \ac{ao}. Then, REL-based and \ac{ael}-based calibration algorithms using \ac{gd} will be detailed, followed by the discussions on cooperative strategies and complexity analysis. The calibration using models M1-M3 can be performed as a subproblem of model M4-based calibration, with the knowledge of beamforming angle (M2) and element pattern (M3) that can be obtained from the chamber.

\label{sec_calibration_algorithm}
\subsection{REL-Based Calibration (AO)}
The idea of \ac{ao} is to optimize one variable while fixing the other one. We start with the most general case (i.e., M4), where both the beamforming matrix $\Wm$ and the channel gain matrix $\Gammam$ need to be estimated jointly. For simplicity, the angle set $\Thetam$ in~\eqref{eq_compact_matrix_form_2} is ignored (i.e., using $\Am$ instead of $\Am(\Thetam)$), and we set $\rho_s=1$ for all the sample angles. 


\subsubsection{Update $\Gammam$ given $\hat\Wm$}
A practical way of initializing \ac{ao} is to 
use the beamforming direction $\bar\Phim$ given by the antenna array manufacturer (see M1 in~\eqref{eq_model1}).
With an estimated $\hat\Wm$, the update on $\Gammam$ is straightforward, which can be formulated as
\begin{equation}
\hat\Gammam = \argmin_{\boldsymbol{\Gamma}} \| \Ym - \hat\Wm^\herm\Am \Gammam \|_\text{F}^2.
\label{eq_update_Gamma}
\end{equation}
Since $\Gammam$ is diagonal, each diagonal element $\gamma_t$ can be updated independently and in parallel as
\begin{equation}
\hat\gamma_t = \argmin_{\gamma} \| \yv_t - \gamma \hat\bv_t \|^2,
\end{equation}
where $\yv_t \in \mathbb{C}^{G \times 1}$ and $\hat\bv_t$ are the $t$-th columns of matrices $\Ym$ and $\hat\Wm\Am$, respectively. The closed-form solution is then given by
\begin{equation}
\hat\gamma_t = \frac{\hat\bv_t^\herm \yv_t}{\hat\bv_t^\herm \hat\bv_t}.
\label{eq_optimal_gamma}
\end{equation}

\subsubsection{Update $\Wm$ given $\hat\Gammam$}


{
With a known $\hat\Gammam$, the update of $\Wm$ under the unit-norm constraint on each column can be formulated as
\begin{equation}
\min_{\|\wv_g\|_2 = 1} 
\| \yv_g - \wv_g^\herm \Am \hat\Gammam \|^2,
\label{eq_constrained_W}
\end{equation}
where $\yv_g$ denotes the $g$-th row of $\Ym$. 
Let $\Bm = \Am \hat\Gammam$, $\Qm = \Bm \Bm^\herm$ (Hermitian psd), and $\bv_g = \Bm \yv_g^\herm$.
The Karush–Kuhn–Tucker (KKT) optimality conditions for~\eqref{eq_constrained_W} (using Wirtinger calculus for complex variables~\cite{brandwood1983complex}, and the trust-region form~\cite{nocedal2006numerical}) yield the following system:
\begin{equation}
(\Qm + \lambda_g \Id) \wv_g = \bv_g, 
\qquad \|\wv_g\| = 1,
\label{eq_kkt_W}
\end{equation}
where $\lambda_g \in \mathbb{R}$ is the Lagrange multiplier.
The function $f(\lambda_g) = \| (\Qm + \lambda_g \Id)^{-1} \bv_g \|$ is strictly decreasing in $\lambda_g$. If the least-squares solution $\wv_g^\dagger=\Qm^\dagger\bv_g$ satisfies $\|\wv_g^\dagger\|_2\le 1$, then it already meets the constraint and the optimum is $\wv_g^\star=\wv_g^\dagger$ (i.e., $\lambda_g^\star=0$). 
Otherwise, its root $\lambda_g^\star$ satisfying $f(\lambda_g) = 1$ can be found efficiently via a one-dimensional bisection search.
Consequently, the resulting vector 
$\wv_g^\star = (\Qm + \lambda_g^\star \Id)^{-1} \bv_g$ satisfies the unit-norm constraint $\|\wv_g^\star\|_2 = 1$ and provides the exact constrained minimizer of~\eqref{eq_constrained_W}. 

}

\subsection{REL-Based Calibration (GD)}

The trust-region formulation in~\eqref{eq_kkt_W} guarantees KKT optimality. However, it incurs a high computational cost since each column $\wv_g$ requires solving a scalar root-finding problem involving the matrix inverse $(\Qm + \lambda_g \Id)^{-1}$. 
This subsection provides a \ac{gd}-based solution for REL-based calibration with a batch size $S_\text{batch}$.
Considering the real-valued loss function defined in~\eqref{eq_loss_1} and its involved complex variables, we develop a gradient-based method to update $\Wm$ and $\gammav$ based on complex gradient operators~\cite{brandwood1983complex} (Theorem 4).
Specifically, the codebook $\Wm$ and the channel gain vector $\gammav$ for each iteration are updated with a learning rate $l_r$ as
\begin{align}
    \Wm^{(i+1)}
    &=\Wm^{(i)}-l_{r}\frac{\partial L_{\mathrm{R}}}{\partial \Wm^*}\Bigg|_{\Wm=\Wm^{(i)}},
    \label{eq_REL_GD_W}
    \\
    \gammav^{(i+1)}
    &=\gammav^{(i)}-l_{r}\frac{\partial L_{\mathrm{R}}}{\partial \gammav^*}\Bigg|_{\gammav=\gammav^{(i)}}.
\end{align}
Note that the calculation of $\Wm^{(i+1)}$ in~\eqref{eq_REL_GD_W} is followed by the normalization of each codeword to fulfill the constraint.  
For REL-based calibration, we need to calculate the gradient term $\partial L_\mathrm{R}/\partial \Wm^*$. Based on~\eqref{eq_REL_loss_function}, we can define the residual 
\begin{equation}
\ev_t = \yv_t - \bv_t = \yv_t - \gamma_t \Wm^\herm \av_t,
\end{equation}
where $\check\bv_t$, $\bv_t$ and $\av_t$ are defined after~\eqref{eq:grad_gamma_dael}. Expanding the squared norm, we get
\begin{equation}
\|\ev_t\|^2 = \yv_t^\herm \yv_t - \yv_t^\herm \gamma_t \Wm^\herm \av_t - \gamma_t^* \av_t^\herm \Wm \yv_t + |\gamma_t|^2 \av_t^\herm \Wm \Wm^\herm \av_t.
\end{equation}
And the gradient of $L_{\mathrm{R}}$ with respect to $\Wm^*$ and $\gamma_t$ can be expressed respectively as
\begin{align}
\frac{\partial L_{\mathrm{R}}}{\partial \Wm^*}
&= \frac{1}{T} \sum_{t=1}^T \left( -\gamma_t \av_t \yv_t^\herm + |\gamma_t|^2 \av_t \av_t^\herm \Wm \right),
\label{eq:gradient_REL}
\\
\frac{\partial L_{\mathrm{R}}}{\partial \gamma_t^*}
&= \frac{1}{T} (-\av_t^\herm \Wm \yv_t + \gamma_t \av_t^\herm \Wm \Wm^\herm \av_t).
\end{align}
Note that the GD-based update on $\gammav$ is not mandatory.
By insterting the updated codebook $\hat\Wm=\Wm- l_{r\text{R}}\frac{\partial L_{\mathrm{R}}}{\partial \Wm^*}$ into~\eqref{eq_update_Gamma} with column normalization, AO-based update on the gain matrix $\Gammam$ can still be performed for REL-based calibration. In the simulation section that follows, we will keep the best-performing method (GD-based update on $\Wm$ only) and show how mini-batch \ac{gd} empirically outperforms \ac{ao} and how the hyperparameters affect the calibration performance.

\subsection{AEL-Based Calibration}


Based on the available estimated $\Wm$, $\Gammam$ (e.g., from REL-based calibration), the current beam response can be calculated as $\Bm = \Wm^\herm \Am(\thetav)\Gammam$. Together with the normalized measurement $\check\Bm$, the iterative optimization can be performed for AEL-based calibration for M4 using the loss function $L_\mathrm{A}$.

To assist derivation, we compute the gradients of the proposed error loss $L_{\text{DA}}$ with respect to the model parameters $\Wm^*$ and $\gamma_t^*$ using Wirtinger calculus, by calculating the derivative of the loss $e_{t,s}$ with respect to the unknown parameters and summing over $T$ samples and $S$ anchors.
The gradients of the proposed error loss $L_{\text{DA}}$ with respect to the model parameters $\Wm^*$ and $\gamma_t^*$ are given by
\begin{align}
\label{eq:grad_w_dael}
\frac{\partial L_{\mathrm{A}}}{\partial \Wm^*} 
& = \frac{\sum_{t,s}\!\left[
\frac{\gamma_t}{q_t}\,e_{t,s}\,\av_t \yv_s^{\herm}
-\frac{|\gamma_t|^2}{q_t^{2}}\,\Re\{e_{t,s}u_{t,s}^*\}\,
\av_t\av_t^{\herm}\Wm
\right]}
{TS},
\\
\label{eq:grad_gamma_dael}
\frac{\partial L_{\mathrm{A}}}{\partial \gamma_t^*}
&=
\frac{j\sum_{s}\rho_{t,s}\,
\Im \{e_{t,s}^* u_{t,s}\}}{\gamma_t^*TS}.
\end{align}
The intermediate variables are defined as $\av_t = \av(\varthetav_t)$, $q_t = |\gamma_t| \|\Wm^\herm \av_t\|$, $p_{t,s} = \gamma_t \av_t^\herm \Wm \yv_s$, $u_{t,s} = \frac{p_{t,s}}{q_t}$, $\check{u}_{t,s} = \frac{\yv_t^\herm \bv_s}{\|\yv_t\|}$, with $\bv_t = \gamma_t \Wm^\herm \av_t$.
The AEL-based calibration is based on the loss function in~\eqref{eq_loss_function_3} and hence there is no closed-form solution for $\gamma_t$, in contrast to the REL-based calibration in~\eqref{eq_optimal_gamma}. The detailed derivation is provided in Appendix~\ref{appendix_A}. 

\subsection{Cooperative Calibration}
\label{sec_cooperative_calibration}
From the calibration models, we can see that the channel gain matrix is only used for beam representation, which aligns the reconstructed beam pattern as closely as the original ground truth pattern. 
When performing localization and sensing, the complex gain will be a function of the target angle and the codebook, as shown in equation~\eqref{eq_maximum_likelihood}. Consequently, the matrix $\Wm$ is the variable of interest for calibration. 

Ideally, the BS can collect all the received signals from several UEs, and calibration can be done centrally based on the ground truth UE state. However, the communication between BS and UEs introduces overhead. Instead, distributed calibration at each UE side can largely reduce the complexity, such as large matrix inversion, and communication overhead. By using a federated learning strategy initialized with a local model broadcasted by the BS as $\Wm^-$, each user $m$ can update the local model with the beam pattern difference $\Delta_{\Wm_{m}}$, and the BS can aggregate the local models as
\begin{equation}
    \Wm^+ = \Wm^- + \sum_m \xi_{m}\Delta_{\Wm_{m}},
    \label{eq_cooperative_calibration}
\end{equation}
where $\xi_{m}$ is the weighting coefficient for the $m$-th local model. While the choice of coefficients (e.g., based on the UE's hardware capability, location, and number of measurements) influences performance, this work focuses on evaluation rather than on their optimization. In the above equation, the nuisance parameter $\Gammam_m$ is not included and the beamforming matrix $\Wm^+$ also needs to be normalized after each update. However, it could be utilized to design the weighting coefficients (e.g., larger coefficients for the model with a larger channel gain).

{
\subsection{Complexity Analysis}
The computational complexity of the proposed calibration methods is analyzed as follows.  
For REL-based calibration with \ac{ao}, the update of $\Gammam$ in~\eqref{eq_update_Gamma} requires matrix multiplication cost as $\mathcal{O}(GNT)$. 
The update of $\Wm$ in~\eqref{eq_constrained_W} requires $\mathcal{O}(GNT + N^{2}T + GN^{2})$ operations for matrix multiplication, followed by eigen decomposition and bisection search with the complexity of $\mathcal{O}(N^{3})$ and $\mathcal{O}(L G N^{2})$, respectively. 
Therefore, the total computational complexity per iteration is 
$\mathcal{O}\big(GNT + N^{2}T + GN^{2} + N^{3} + L G N^{2}\big)$.
For REL-based calibration using \ac{gd} in~\eqref{eq:gradient_REL}, the gradient evaluation can be efficiently implemented via rank-one updates, resulting in a per-sample cost of $\mathcal{O}(GN)$ and hence $\mathcal{O}(GNT_b)$ per iteration with batch size $T_b$.
For AEL-based calibration in~\eqref{eq:grad_w_dael} and~\eqref{eq:grad_gamma_dael}, evaluating the gradient requires computing $a_{s}^{\herm}\Wm$ once per sample at cost $\mathcal{O}(NG)$, and then accumulating across $T_b$ samples and $S$ anchors, resulting a complexity of $\mathcal{O}(GNT_bS)$.

In summary, REL-based calibration using \ac{gd} scales linearly with the batch size and the number of antennas and anchors, providing a flexible and robust solution towards calibration. In addition, we notice that the AEL-based algorithm has a higher complexity than the REL-based calibration, depending on the anchor size $S$. 
}



\begin{figure}[!t]
\centering
\centerline{\includegraphics[width=1\linewidth]{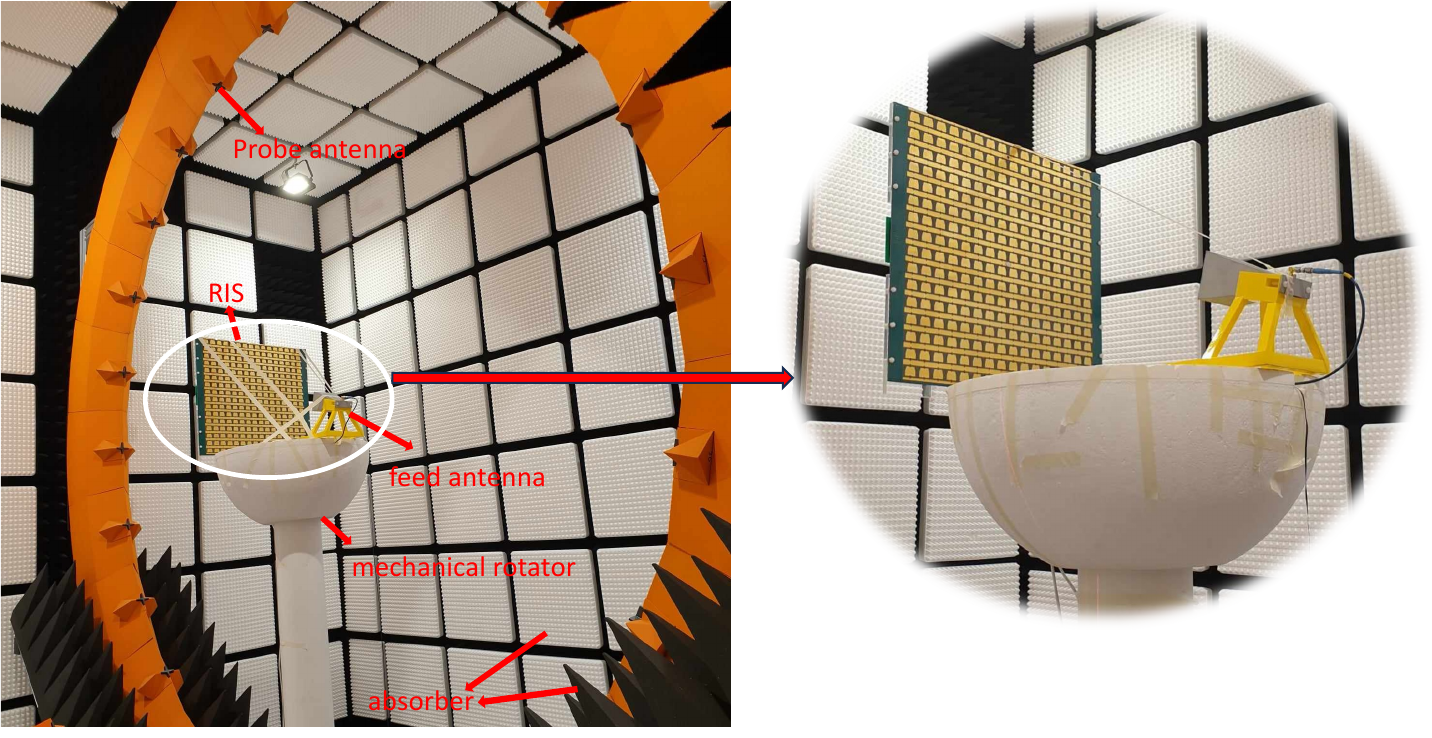}}
\vspace{-0.1cm}
\caption{The photo of the RIS pattern measurements in the anechoic chamber.}
\label{fig_RISmea_setup}
\vspace{-0.5cm}
\end{figure}

\section{Simulation and Experimental Validation}\label{sec:validation}
We start with the measurement data from the chamber, which serves two purposes: (i) to provide ideal measured data to evaluate different beam pattern models, and (ii) to benchmark the developed calibration algorithms in more practical settings (e.g., different transmit powers). However, the proposed cooperative calibration can also be applied in practical scenarios. In addition, both 2D and 3D beam pattern calibration results will be presented, and different calibration scenarios will also be evaluated.

\begin{figure*}[h]
\centering
\begin{minipage}[b]{0.32\linewidth}
  \centering
\centerline{\includegraphics[width=0.99\linewidth]{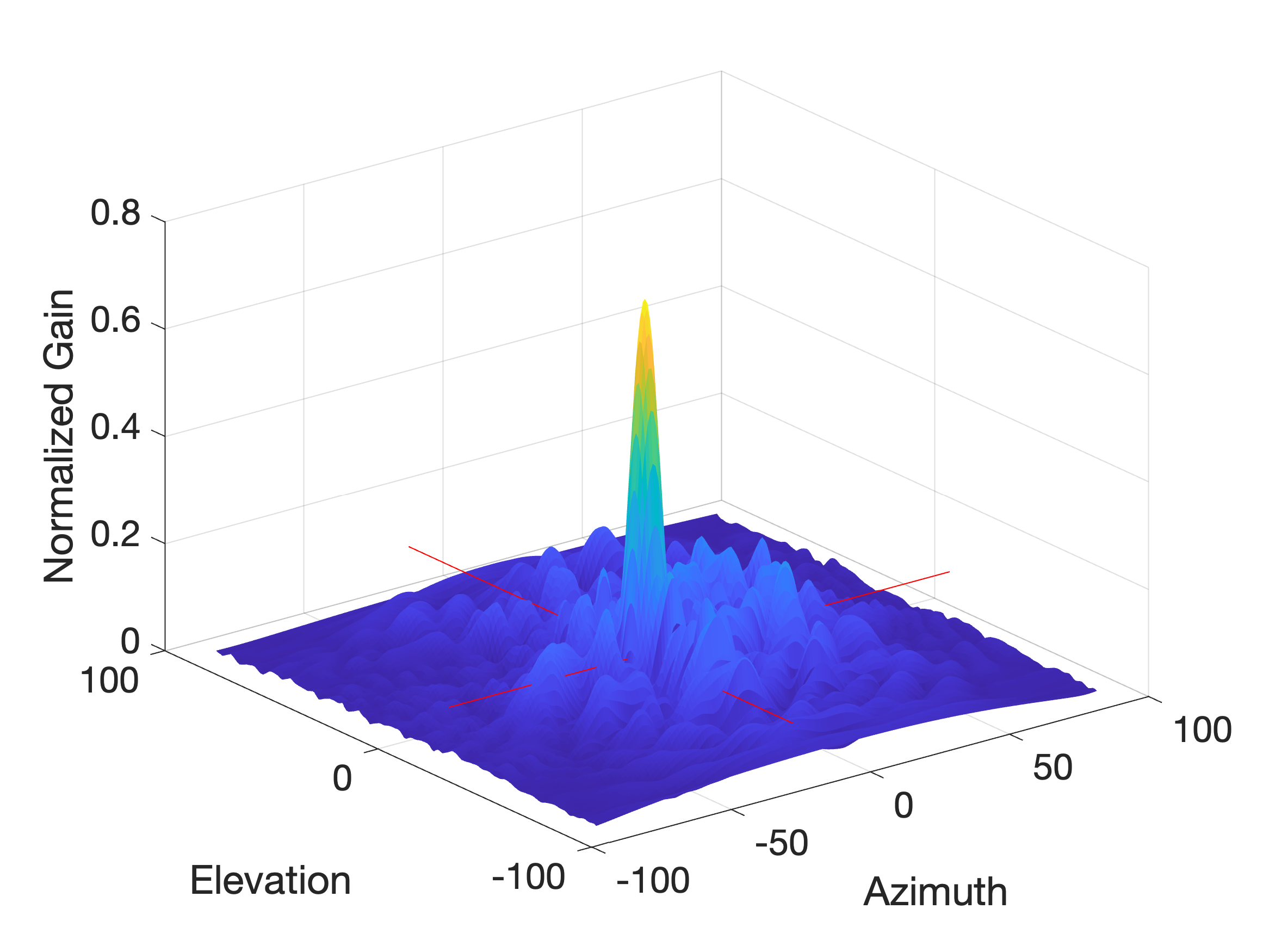}}
  \centerline{\small{(a)}}
\end{minipage}
\begin{minipage}[b]{0.32\linewidth}
\centering
    \include{Figures_tikz/fig03b}   
    \vspace{-1cm}
  \centerline{\small{(b)}} 
\end{minipage}
\begin{minipage}[b]{0.32\linewidth}
\centering
\centerline{\includegraphics[width=0.99\linewidth]{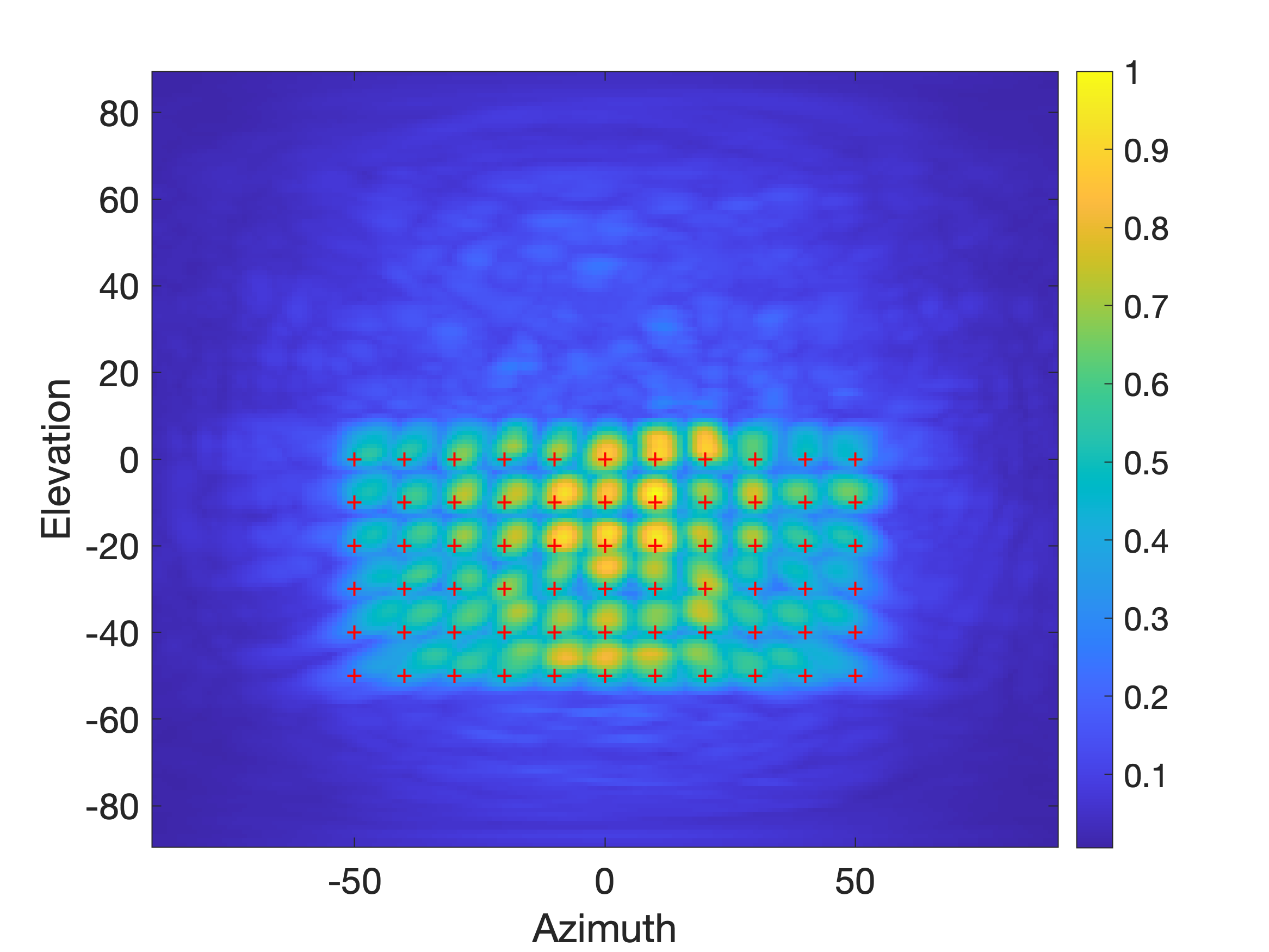}}
\centerline{\small{(c)}} 
\end{minipage}
\caption{{Visualization of beam pattern measurements collected from the chamber: (a) 3D beam pattern for scanning angle at $[-20^\circ, -20^\circ]$; (b) Visualization of 2D beam patterns ($11$ codewords) when elevation angle is $0^\circ$}; (c) Visualization of 3D beam patterns ($6 \times 11$ codewords, and the red cross indicates ideal beamforming angle).}
\label{fig_beam_pattern_visualization}
\vspace{-0.5cm}
\end{figure*}

\subsection{System Setup}


To obtain the ground truth calibration data, we measured the reflected beam pattern of a real-world RIS plate\footnote{The proposed calibration models can be implemented in both analog arrays and passive RISs. In this work, the transmitter and the RIS board have a fixed position and are treated as an analog array.} in the anechoic chamber, as shown in Fig.~\ref{fig_RISmea_setup}. 
The specifications of the RIS board under test are summarized in Table~\ref{table_RIS_parameters}, and more details on its design can be found in \cite{li2023design}. In each beam, RIS coefficients can be programmed to assign the phase shift values to maximize the energy in this direction, and the corresponding beam pattern requires calibration. 

Here, we measured $G = 6 \times 11 = 66$ beam patterns with its target boresight direction pointing at $\bar\Phim = [\bar \phi^{\mathrm{az}},\bar \phi^{\mathrm{el}}]^\top$ where $\bar \phi^{\mathrm{az}} \in \{-50^\circ, -40^\circ, \ldots, 50^\circ\}$, $\bar \phi^{\mathrm{el}} \in \{-50^\circ, -40^\circ, \ldots, 0^\circ\}$, respectively. The SNR of the measurements in the chamber was around $35$ dB.
Note that the boresight direction of the practical RIS beams can be different from the target angles, which is one of the key motivations for beam calibration. The resulting beam patterns were measured over a hemispherical range, with the azimuth angle varying from $-90^\circ$ to $90^\circ$ and the elevation angle from $-90^\circ$ to $90^\circ$, both sampled at $1^\circ$ intervals, resulting in a measurement dataset $\bar\Bm_\text{3D} \in \mathbb{C}^{66 \times 181 \times 181}$. For 2D beam pattern calibration, we set the elevation angle as $0^\circ$ and use a subset of the measurement data as $\bar\Bm_\text{2D} \in \mathbb{C}^{11\times 181}$.
The measurement data is used as the ground truth and for generating synthetic measurement data for calibration performance evaluation in different scenarios.

\begin{table}[h]
    \centering
    \caption{RIS Specifications and Calibration Parameters}
    \begin{tabular}{l | c | c}
        \toprule
        \textbf{Parameters} & \textbf{2D} & \textbf{3D}\\
        \midrule
        Operation frequency & \multicolumn{2}{c}{5 GHz} \\
        Phase-shifter resolution & \multicolumn{2}{c}{2-bit (0$^\circ$, 90$^\circ$, 180$^\circ$ and 270$^\circ$)}
        \\
        Polarization & \multicolumn{2}{c}{linear polarization} \\
        $G$ (beams) & 11 & 66 
        \\
        $N$ (antennas) & 16 & 256
        \\        
        \midrule
        $T$ (sample angles) & 561 & 12831
        \\
        Sample angle range (azi.) & $[-70^\circ, 70^\circ]$ & $[-70^\circ, 70^\circ]$\\
        Sample angle range (ele.) & - & $[-70^\circ, 20^\circ]$
        \\
        $S$ (evaluation angles) & 81 & 119
        \\
        Anchor angle range (azi.) & $[-40^\circ, 40^\circ]$ & $[-40^\circ, 40^\circ]$\\
        Anchor angle range (ele.) & - & $[-40^\circ, 10^\circ]$
        \\
        $M$ (cooperative users) & 3 & -
        \\
        $l_{r\text{R}}$: REL learning rate & 0.02 & 0.1
        \\
        $l_{r\text{A}}$: REL learning rate & 0.03 & 0.2
        \\
        \bottomrule
    \end{tabular}
    \label{table_RIS_parameters}
\end{table}

\subsection{Visualization of Measurement Data}

The visualization of a specific 3D codeword (beamforming direction $[0^\circ, 0^\circ]^\top$) can be found in Fig.~\ref{fig_beam_pattern_visualization} (a). 
When considering 2D beam pattern calibration, the beam patterns of $11$ codewords are shown in Fig.~\ref{fig_beam_pattern_visualization} (b).
For better visualization of the whole codebook in 3D calibration, we extract the largest gain (e.g., getting the maximum amplitude in the first dimension of the reshaped measurement dataset $\bar{{\Bm}}_\text{3D}$) at certain angles as shown in Fig.~\ref{fig_beam_pattern_visualization} (c). We can clearly see the offset between the main lobe direction and the expected beamforming direction $\bar\Phim$ (red cross markers), necessitating the need for calibration.


\begin{table}[h]
    \centering
    \caption{Beam Pattern Calibration Results (in terms of beam similarity $S_\mathrm{R}$, angle error $E_\mathrm{A}$, and codebook gain loss $E_\mathrm{C}$)}
    \begin{tabular}{l l l l l}
        \toprule
        \textbf{} & \textbf{Model} 
        & \textbf{$S_\mathrm{R}$}[\%] & \textbf{$E_\mathrm{A}$}[$^\circ$] & \textbf{$E_\mathrm{C}$}[dB] \\
        \midrule
        \multirow{5}{*}{2D} 
        & M1 (uncalibrated) & 80.60 & 1.0123 & -1.4019 \\
        & M2 & 80.62 & 0.7926 & -1.2367 \\
        & M3 & 80.40 & 2.5284 & -1.5944 \\
        & M4 ($L_{\mathrm{R}}$-AO) 
        & 96.45 & 0.1654  & -1.1640\\
        & M4 ($L_{\mathrm{R}}$-GD) 
        & \textbf{96.81} & 0.1506 & \textbf{-1.1636} \\
        & M4 ($L_{\mathrm{A}}$) 
        & 95.81 & \textbf{0.1086} & \textbf{-1.1636} \\
        \midrule
        \multirow{5}{*}{3D} 
        & M1 (uncalibrated) & 63.72 & 5.1874 & -2.7750 \\
        & M2 & 63.30 & 3.5235 & -1.6316 \\
        & M3 & 78.79 & 2.3497 & -3.2833\\
        & M4 ($L_{\mathrm{R}}$-AO) & 95.65 & 0.9559 & \textbf{-1.5236}\\
        & M4 ($L_{\mathrm{R}}$-GD) & \textbf{97.30} & 0.8906 & -1.5250 \\
        & M4 ($L_{\mathrm{A}}$) & 93.93 & \textbf{0.8556} & {-1.5241}\\
        \bottomrule
    \end{tabular}
    \label{table_calibration_results}
\end{table}


\subsection{Calibration Performance Evaluation}
The calibration methods described in this work are implemented in 2D and 3D scenarios. The comparison of different calibration methods on various performance metrics and models is shown in Table~\ref{table_calibration_results}. We notice that the calibration model M4, which considers estimating both the beamforming matrix $\Wm$ and channel gain matrix $\Gammam$, achieves the best performance. Furthermore, the implementation of the loss function $L_\text{DA}$ further improves AEL, showing the effectiveness of the proposed loss function. We also noticed that the calibration error of 3D beam patterns is much larger than the 2D benchmark, and the improvement using AEL is limited. This is due to the higher dimension of calibration parameters that produces a large number of local minima, which affect the gradient-based calibration processes and lead to suboptimal solutions. In summary, the calibration method in this work can improve the angle estimation performance from $1.01^\circ / 5.19^\circ$ to $0.11^\circ / 0.86^\circ$ in 2D/3D scenarios, respectively.

\begin{figure}[t]
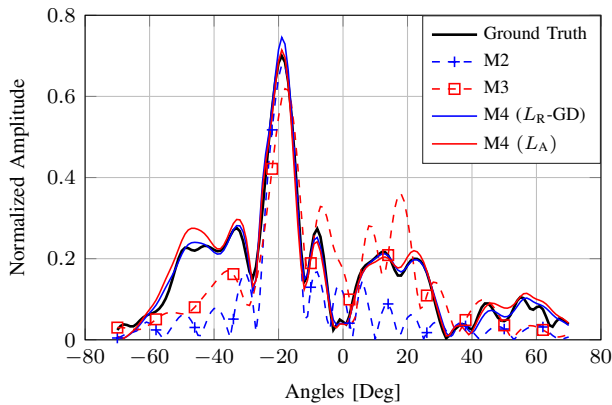

\centering
\include{Figures_tikz/fig04}
\vspace{-0.8 cm}
\vspace{-0.1cm}
\caption{Visualization of the reconstructed 2D beam patterns ($\bar\varphi=-20^\circ$).}
\label{fig_calibrated_beams_2d}
\end{figure}

\begin{figure}[t]
\centering
\begin{minipage}[b]{0.48\linewidth}
  \centering
\centerline{\includegraphics[width=1.09\linewidth]{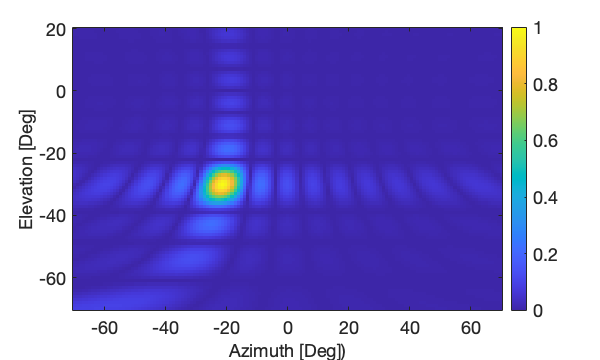}}
  \centerline{\small{(a) Ground truth}} \medskip
\end{minipage}
\vspace{-0.2cm}
\begin{minipage}[b]{0.48\linewidth}
\centering
\centerline{\includegraphics[width=1.09\linewidth]{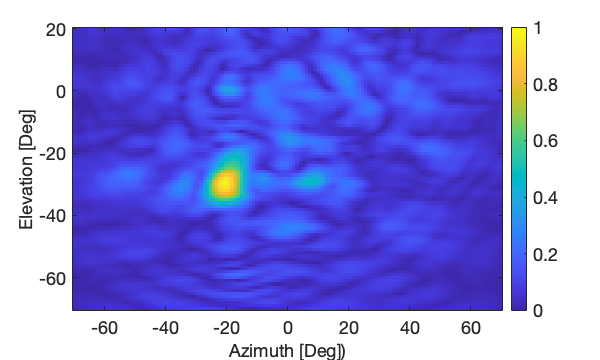}}
  \centerline{\small{(b) Ideal beam pattern}} \medskip
\end{minipage}

\begin{minipage}[b]{0.48\linewidth}
  \centering
\centerline{\includegraphics[width=1.09\linewidth]{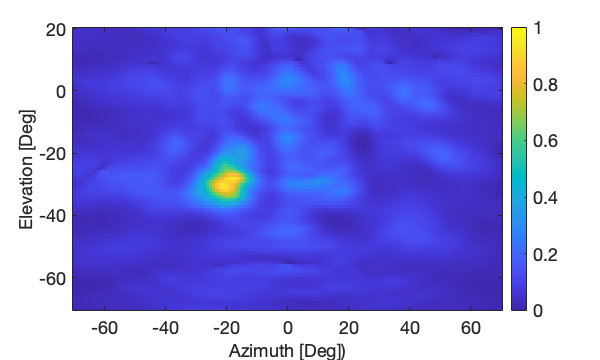}}
  \centerline{\small{(c) M4 ($L_\text{R}$-GD)}} \medskip
\end{minipage}
\vspace{-0.2cm}
\begin{minipage}[b]{0.48\linewidth}
\centering
\centerline{\includegraphics[width=1.09\linewidth]{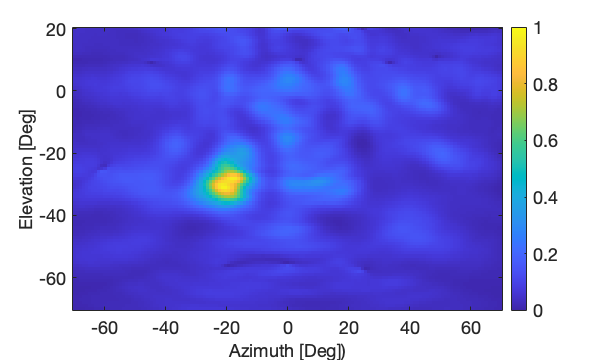}}
  \centerline{\small{(d) M4 ($L_\text{DA}$)}} \medskip
\end{minipage}
\caption{{Visualization of the reconstructed 3D beam patterns.}}
\label{fig_calibrated_beams_3d}
\end{figure}

\begin{figure}[h]
\centering
\begin{minipage}[b]{0.9\linewidth}
  \centering
\centerline{\includegraphics[width=0.8\linewidth]{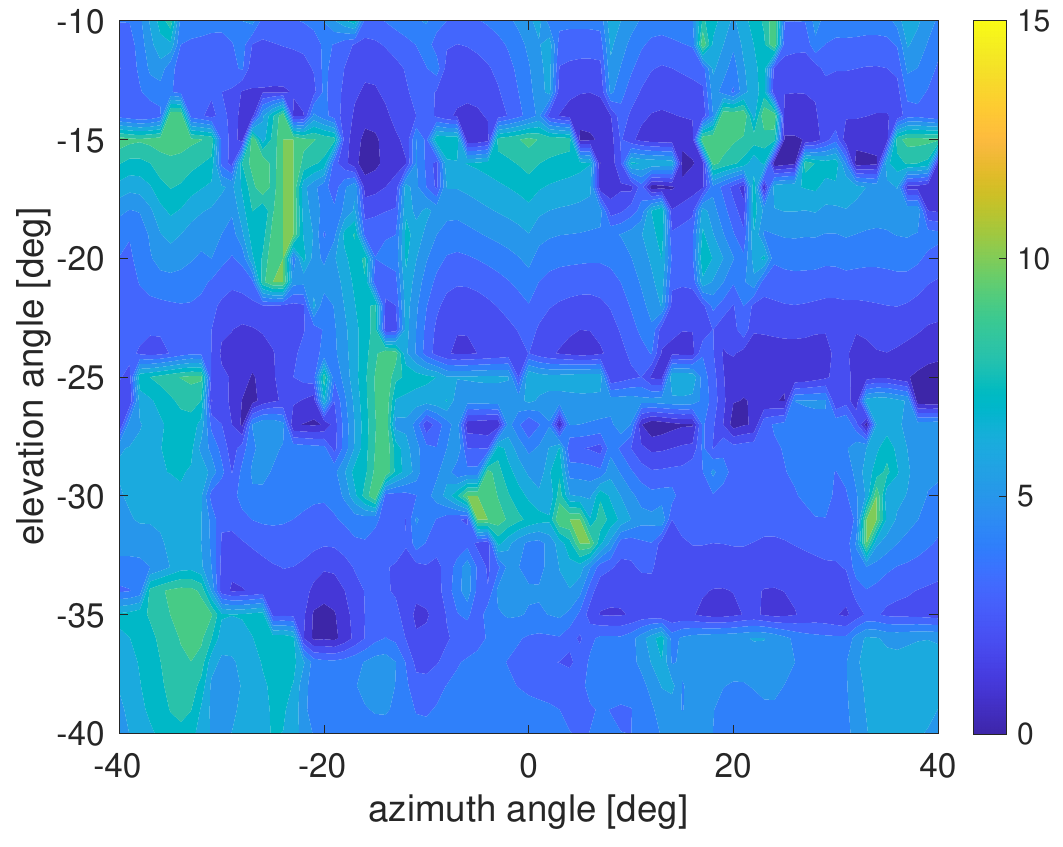}}
  \centerline{\small{(a) M1}} \medskip
  \vspace{-0.1cm}
\end{minipage}
\vspace{-0.2cm}
\begin{minipage}[b]{0.9\linewidth}
\centering
\centerline{\includegraphics[width=0.8\linewidth]{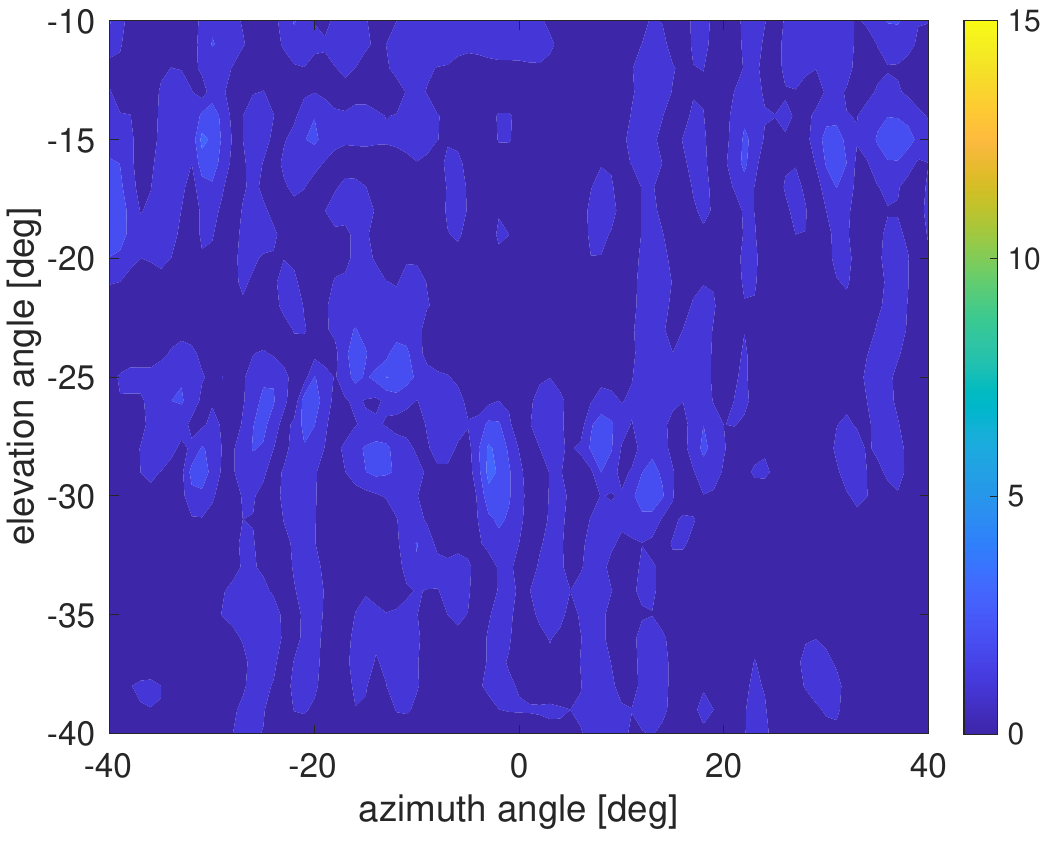}}
  \centerline{\small{(b) M4 ($L_{\mathrm{A}}$)}} \medskip
  \vspace{-0.2cm}
\end{minipage}
\caption{Visualization of angle estimation errors before and after calibration.}
\label{fig_angle_estiation_error_heatmap}
\vspace{-0.3cm}
\end{figure}

The visualization of one specific calibrated 2D beam patterns (beamforming angle $\bar\phi^{\text{az}} = -20^\circ$) is shown in Fig.~\ref{fig_calibrated_beams_2d}. 
Here, we omit M1 and M4 ($L_R$) due to the better performance of M2 and M4 ($L_\text{R}$-GD), respectively. 
Despite the superior performance of $L_\text{R}$-based calibrations over $L_\text{DA}$ in REL, the reconstructed beam pattern based on $L_\text{DA}$ is closer to the ground truth beam pattern in the main lobe, which is also reflected from other beam patterns that are not visualized.
The visualization of calibrated 3D beam patterns (beamforming angle $\bar\varphiv = [-20^\circ, -20^\circ]^\top$) is shown in Fig.~\ref{fig_calibrated_beams_3d}. The calibrated beam patterns in Fig.~\ref{fig_calibrated_beams_3d}(c) and Fig.~\ref{fig_calibrated_beams_3d}(d) can capture the features of ground truth much better than the uncalibrated one in Fig.~\ref{fig_calibrated_beams_3d}(b). It is also seen that there are some outliers in AEL-based calibration on edge angles, which is due to the selection of evaluation angles.
{The heatmap of angle estimation bias using the calibrated beam pattern is shown in Fig.~\ref{fig_angle_estiation_error_heatmap}, where the error level in all areas of interest can be largely reduced using the calibrated beam pattern.}


\subsection{Evaluation of Calibration Parameters}
\begin{figure}[h]
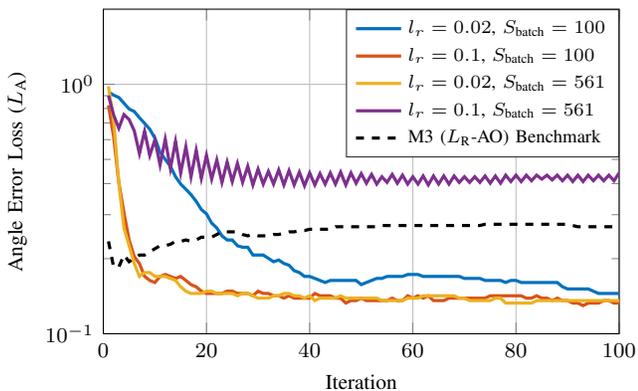

\centering
\include{Figures_tikz/fig07}
\vspace{-0.8 cm}
\vspace{-0.1cm}
\caption{Evaluation of different parameters using REL-based calibration (\ac{gd}).}
\label{fig_hyperparameters_gradient_descent}
\vspace{-0.2cm}
\end{figure}

Due to the high calculation complexity using 3D beam data, we next focus on 2D beam pattern calibration and evaluate the impact of different parameters on calibration model M4. 
We first assess the performance improvement using \ac{gd}-based method compared with AO in REL-based calibration, as shown in Fig.~\ref{fig_hyperparameters_gradient_descent}. 
It is shown that the AO-based benchmark cannot converge well even with a sufficient number of iterations.
In contrast, the \ac{gd}-based method performs better with sufficient iterations and appropriate parameters. Specifically, a smaller mini-batch size requires a larger learning rate compared with processing all the measurement data at the same time. However, when the learning rate is too high, performance will be affected with oscillating loss, as shown in the curve with $l_r=0.1$ with the whole batch processing. It is foreseen that more advanced optimizers can be adopted for better performance, which is beyond the scope of this work.

\begin{figure}[h]
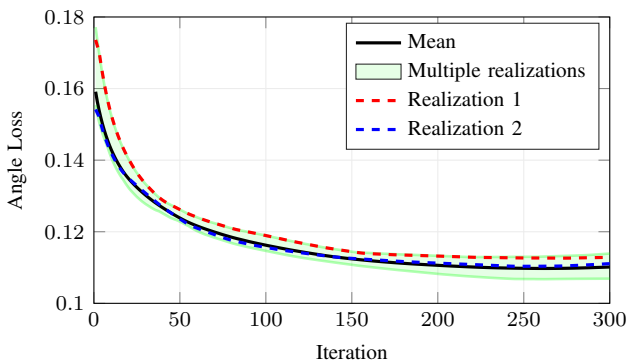

\centering
\include{Figures_tikz/fig08}
\vspace{-0.5 cm}
\vspace{-0.5cm}
\caption{AEL-based calibration with different initial codebooks $\Wm$ obtained using REL-based calibration.}
\label{fig_different_initial_points}
\end{figure}

With a calibrated beam based on $L_\text{R}$, AEL-based calibration can be performed. The evaluation of 20 different initial beamforming matrices $\Wm$ is shown in Fig.~\ref{fig_different_initial_points}.
Specifically, multiple realizations are visualized in the green area, with two specific realizations and mean loss plotted in dashed and solid curves. 
It is seen that the initial point directly affects calibration performance. Nevertheless, the proposed AEL-based calibration loss function can stably decrease the angle error in general.

{For more practical consideration, we evaluate the impact of measurement noise on calibration. 
In this simulation, the BS is located at $[0, 0]~\text{m}$, and the UE moves from $[10, -30]~\text{m}$ to $[10, 30]~\text{m}$ along the line $x = 10~\text{m}$, with samples taken every $0.2\,\text{m}$ along the $y$-axis. Three scatterers are evenly spaced along the $y$-range $[-30, 30]~\text{m}$ at $x = 10$, and each scatterer is surrounded by 5 scattering points randomly distributed within a 1-meter radius circle centered at the scatterer.
 With different transmit power, the calibration performance (using angle loss) of M4-AEL is shown in Fig.~\ref{fig_noisy_mpc_observation} (a). Benchmarked against the scenario without noise and MPCs, the angle loss in noise-free calibration with MPCs increases the angle error from $0.21^\circ$ to $0.33^\circ$. The blue, yellow, and red curves combine the effects of received signal noise and the MPCs, showing that the calibration performance will be affected by strong MPCs (e.g., RCS = 10). As for the effect of signal noise, it can be mitigated by increasing transmit power. We further investigate different calibration scenarios at the transmit power of $15$ dBm. The CDF of angle estimation error is shown in
Fig.~\ref{fig_noisy_mpc_observation} (b), highlighting a large gap caused by MPCs in the range of $0.2\le \epsilon \le 1.2$.

}


\begin{figure}[t]
\centering
\begin{minipage}[b]{0.98\linewidth}
  \centering
%
%
\definecolor{mycolor1}{rgb}{0.00000,0.44700,0.74100}%
\definecolor{mycolor2}{rgb}{0.85000,0.32500,0.09800}%
\definecolor{mycolor3}{rgb}{0.92900,0.69400,0.12500}%
\definecolor{mycolor4}{rgb}{0.46600,0.67400,0.18800}%
\begin{tikzpicture}

\begin{axis}[%
width=2.7in,
height=1.7in,
at={(0in,0in)},
scale only axis,
xmin=-5,
xmax=30,
xlabel style={font=\color{white!15!black},font=\footnotesize},
xlabel={Transmit Power [dBm]},
ymin=0.2,
ymax=1,
ylabel style={font=\color{white!15!black},font=\footnotesize},
ylabel={Angle Loss},
tick label style={font=\footnotesize},
axis background/.style={fill=white},
xmajorgrids,
ymajorgrids,
legend style={legend cell align=left, align=left, font=\footnotesize, draw=white!15!black}
]

\addplot [color=mycolor2, line width=1.0pt, mark size=2.0pt, mark=asterisk, mark options={solid, mycolor2}]
  table[row sep=crcr]{%
-5	0.923435984347398\\
2	0.425676511334258\\
9	0.402659940790415\\
16	0.359130161669777\\
23	0.354616446616433\\
30	0.350340733088127\\
};
\addlegendentry{With noise and MPCs (RCS = 10)}

\addplot [color=mycolor3, line width=1.0pt, mark size=2.0pt, mark=diamond, mark options={solid, mycolor3}]
  table[row sep=crcr]{%
-5	0.837677682394448\\
2	0.308172516888193\\
9	0.262340430224781\\
16	0.242636463450275\\
23	0.234649625026932\\
30	0.219354724665958\\
};
\addlegendentry{With noise and MPCs (RCS = 2)}

\addplot [color=mycolor4, line width=1.0pt, mark size=2.0pt, mark=square, mark options={solid, mycolor4}]
  table[row sep=crcr]{%
-5	0.331835429771345\\
2	0.331835429771345\\
9	0.331835429771345\\
16	0.331835429771345\\
23	0.331835429771345\\
30	0.331835429771345\\
};
\addlegendentry{With only MPCs (RCS = 10)}

\addplot [color=mycolor1, line width=1.0pt, mark size=2.0pt, mark=o, mark options={solid, mycolor1}]
  table[row sep=crcr]{%
-5	0.648521745292968\\
2	0.277020698799754\\
9	0.234610627575659\\
16	0.221150188826856\\
23	0.214427132753833\\
30	0.2157240783061\\
};
\addlegendentry{With only noise}

\addplot [color=black, line width=1.0pt]
  table[row sep=crcr]{%
-5	0.213346703177649\\
2	0.213346703177649\\
9	0.213346703177649\\
16	0.213346703177649\\
23	0.213346703177649\\
30	0.213346703177649\\
};
\addlegendentry{Without noise and MPCs}

\end{axis}

\end{tikzpicture}%
     \vspace{-.8 cm}
  \centerline{\small{(a)} } \medskip
\end{minipage}
\vspace{0.2cm}
\begin{minipage}[b]{0.98\linewidth}
\centering
     \include{Figures_tikz/fig09b}    
     \vspace{-.8 cm}
  \centerline{\small{(b)}} \medskip
  \vspace{-0.3cm}
\end{minipage}
\vspace{-0.2cm}
\caption{Evaluation of the impacts of MPCs and transmit power on calibration performance. (a) Angle loss vs. transmit power; (b) CDF of the angle loss under different scenarios.}
\label{fig_noisy_mpc_observation}
\vspace{-0.5cm}
\end{figure}

\subsection{Cooperative Calibration}
\begin{figure}[h]
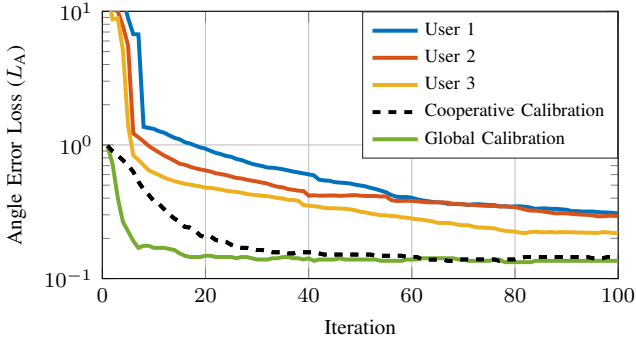

\centering
\include{Figures_tikz/fig10}
\vspace{-0.8 cm}
\vspace{-0.1cm}
\caption{Angle error loss evaluation for local calibration, cooperative calibration, and global calibration.}
\label{fig_local_vs_global}
\vspace{-0.1cm}
\end{figure}

We further evaluate the performance using a cooperative calibration strategy, where $M=3$ UEs process the local dataset to obtain a local beam pattern, which is then used for global fusion. The total dataset $T = 561$ is randomly split into 3 subsets with $T_1 = 100$, $T_2 = 200$, and  $T_3 = 261$. The AELs for all the users using local calibrated results are shown in~Fig.~\ref{fig_local_vs_global}, benchmarked by the global calibration with all of the measurements (solid green curve). The cooperative calibration result is shown in the dashed black curve. For each point of the fused results, it reflects merging the calibrated local parameters after $i$ iterations $\Wm_{m, i}$ with equal coefficients as described in~\eqref{eq_cooperative_calibration}. It is shown that with a sufficient number of iterations, cooperative calibration can achieve comparable performance as the global calibration strategy with a much reduced communication overhead when the measurements are large (reduced to $M\times  G\times N$ from $M\times  G\times T$ in terms of complex entries). Note that the calibration performance is also decided by the coverage of the data, and calibration strategies that consider the UE trajectory can be discussed in future work.

\begin{figure}[h]
\centering
\vspace{-0.2cm}
\centerline{\includegraphics[width=0.9\linewidth]{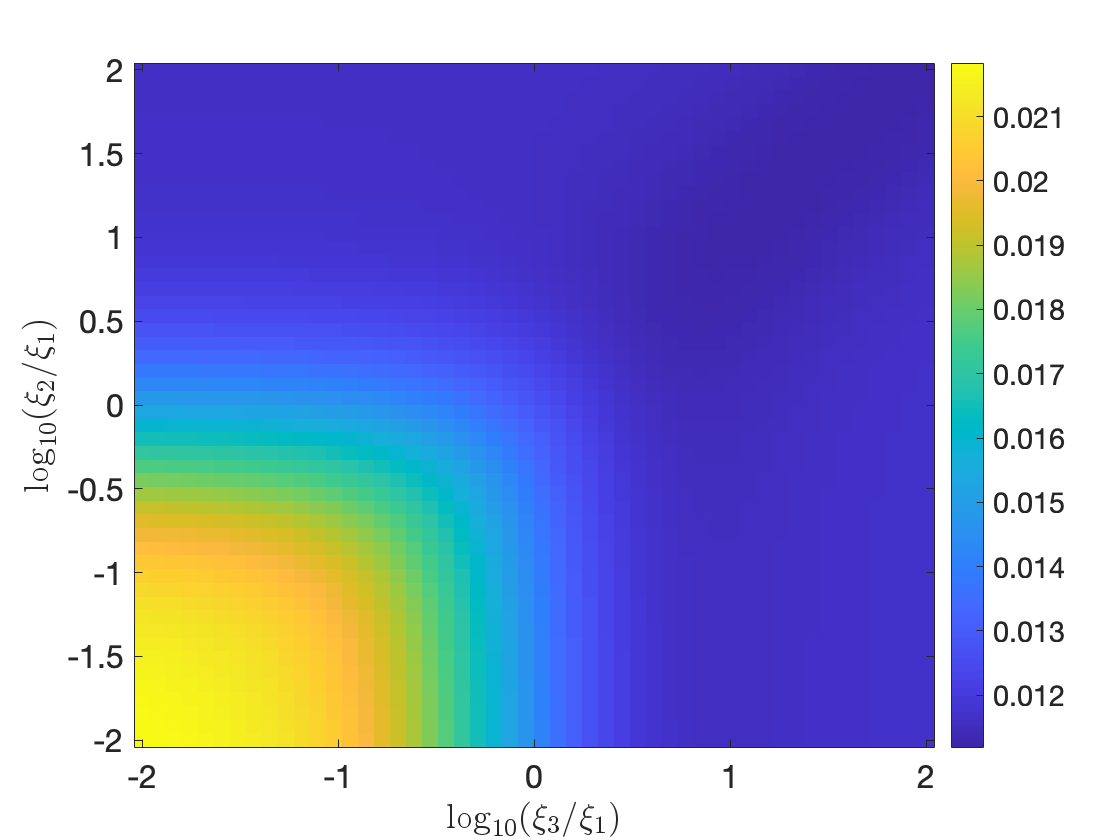}}
\vspace{-0.1cm}
\caption{Weights selection for global fusion.}
\label{fig_weights_for_global}
\end{figure}

The evaluation of different weighting factors is shown in Fig.~\ref{fig_weights_for_global}. Instead of using equal weights as $\xi_1 = \xi_2 = \xi_3 = 1/3$ in Fig.~\ref{fig_local_vs_global}, we take $\xi_1$ as the reference and choose $\xi_2$ and $\xi_3$ from $\xi_1/100$ to $100\xi_1$ and visualize the AEL.
As can be inferred from the figure, large values of $\xi_2$ and $\xi_3$ result in a better calibration performance (shown in green). This is also verified from Fig.~\ref{fig_local_vs_global} that UE 1 has the worst performance, and should be allocated a small weight. However, these results only demonstrate the effectiveness of the cooperative strategy. 
More optimized weighting factor values can be investigated, e.g., based on a Bayesian framework.

\section{Conclusion}
This work presents a comprehensive framework for beam pattern calibration tailored to ISAC systems, prioritizing angular sensing accuracy over traditional pattern similarity metrics. A novel sensing-oriented metric based on angle estimation bias, derived through minimization of the KL divergence, is introduced to evaluate calibration quality. To enable gradient-based optimization, a corresponding differentiable loss function is developed, supporting scalable and effective calibration across multiple user devices. Experimental validation using measured beam patterns from an anechoic chamber setup confirms the practical effectiveness of the proposed models and algorithms. Furthermore, the cooperative calibration strategy demonstrates that distributed updates from multiple UEs can achieve performance on par with centralized methods while significantly reducing communication overhead. These results establish a solid foundation for robust and adaptive beam calibration in future ISAC-enabled 6G systems. Looking ahead, future work should consider calibration algorithms that account for multipath propagation and non-ideal, frequency-dependent beam patterns in wideband systems.

\begin{appendices}
\section{Gradient Derivations for $L_{\mathrm{A}}$}
\label{appendix_A}
Recall the loss function $L_{\mathrm{A}}$ defined as
\begin{equation}
L_{\mathrm{A}} 
=
\frac{\sum_{t,s} \rho_{t,s} \left| e_{t,s} \right|^2}
{\sum_{t,s} \rho_{t,s}}
= 
\frac{\sum_{t,s} \rho_{t,s} \left| u_{t,s} - \check{u}_{t,s} \right|^2}
{\sum_{t,s} \rho_{t,s}},
\end{equation}
where $u_{t,s} = {\bv_t^\herm \yv_s}/{\|\bv_t\|}$, $\check{u}_{t,s} = {\yv_t^\herm \yv_s}/{\|\yv_t\|}$, $\bv_t = \gamma_t \Wm^\herm \av_t$, $\av_t = \av(\varthetav_t)$.
By using Wirtinger calculus, the gradient with respect to the conjugate of the matrix variable $\Wm^*$ can be calculated as
\begin{equation}
\frac{\partial L_{\mathrm{A}}}{\partial \Wm^*} = 
\frac{\sum_{t,s} \rho_{t,s} 
[(u_{t,s} - \check{u}_{t,s})^* \frac{\partial u_{t,s}}{\partial \Wm^*}
+(u_{t,s} - \check{u}_{t,s}) \frac{\partial u^*_{t,s}}{\partial \Wm^*}]}
{\sum_{t,s} \rho_{t,s}} .
\end{equation}
We further define $u_{t,s} = p_{t,s}/q_t$ with $p_{t,s} = \gamma_t^* \av_t^\herm \Wm \yv_s$ and $q_t = |\gamma_t| \|\Wm^\herm \av_t\|$. Using the quotient rule yields
\begin{align}
\frac{\partial u_{t,s}}{\partial \Wm^*} & = 
\frac{q_t \frac{\partial p_{t,s}}{\partial \Wm^*} - p_{t,s} \frac{\partial q_t}{\partial \Wm^*}}{q_t^2},
\\ 
\frac{\partial u_{t,s}^*}{\partial \Wm^*} & = 
\frac{q_t \frac{\partial p_{t,s}^*}{\partial \Wm^*} - p_{t,s}^* \frac{\partial q_t}{\partial \Wm^*}}{q_t^2}
\end{align}
where
\begin{align}
\frac{\partial p_{t,s}}{\partial \Wm^*} = 0, 
\ \ 
\frac{\partial p_{t,s}^*}{\partial \Wm^*} = \gamma_t \av_t \yv_s^\herm, 
\ \ 
\frac{\partial q_t}{\partial \Wm^*} = \frac{|\gamma_t|^2 \av_t \av_t^\herm \Wm}{2q_t}.
\end{align}
{And the final expression is given by
\begin{equation}
\begin{split}
\frac{\partial L_{\mathrm{A}}}{\partial \Wm^*}
& = \frac{\sum_{t,s} \rho_{t,s}(u_{t,s} - \check{u}_{t,s})^* \left( - \frac{u_{t,s} |\gamma_t|^2 \av_t \av_t^\herm \Wm}{2 q_t^2} \right)}{\sum_{t,s} \rho_{t,s}}
\\
& \!\!\! + \frac{\sum_{t,s} \rho_{t,s}(u_{t,s} - \check{u}_{t,s}) \left( \frac{\gamma_t \av_t \yv_s^\herm}{q_t} - \frac{u_{t,s}^*|\gamma_t|^2 \av_t \av_t^\herm \Wm}{2q_t^2} \right)}{\sum_{t,s} \rho_{t,s}}
\\
& \!\!\! = \frac{\sum_{t,s}\rho_{t,s}\!\left[
\frac{\gamma_t}{q_t}\,e_{t,s}\,\av_t \yv_s^{\herm}
-\frac{|\gamma_t|^2}{q_t^{2}}\,\Re\{e_{t,s}u_{t,s}^*\}\,
\av_t\av_t^{\herm}\Wm
\right]}
{\sum_{t,s}\rho_{t,s}}
.
\end{split}
\end{equation}}

For the scalar variable $\gamma_t^*$, we compute the derivative as
\begin{equation}
\frac{\partial L_{\mathrm{A}}}{\partial \gamma_t^*} =
\frac{\sum_{s=1}^S \rho_{t,s} (e_{t,s}^* \frac{\partial u_{t,s}}{\partial \gamma_t^*} + e_{t,s} \frac{\partial u_{t,s}^*}{\partial \gamma_t^*})}{\sum_{t,s} \rho_{t,s}} .
\end{equation}
Based on $\frac{\partial p_{t,s}}{\gamma_t^*} = \frac{p_{t,s}}{\gamma_t^*}$, $\frac{\partial p_{t,s}^*}{\gamma_t^*} = 0$ and $\frac{\partial q_t}{\gamma_t^*} = \frac{q_t}{2\gamma_t^*}$, we can have 
\begin{align}
\frac{\partial u_{t,s}}{\partial \gamma_t^*}
&= \frac{q_t\,\frac{\partial p_{t,s}}{\partial \gamma_t^*}
      - p_{t,s}\,\frac{\partial q_t}{\partial \gamma_t^*}}{q_t^2}
= \frac{p_{t,s}}{2\gamma_t^* q_t}
= \frac{1}{2}\,\frac{u_{t,s}}{\gamma_t^*},\\[4pt]
\frac{\partial u_{t,s}^*}{\partial \gamma_t^*}
&= -\,\frac{p_{t,s}^*}{2\gamma_t^* q_t}
= -\,\frac{1}{2}\,\frac{u_{t,s}^*}{\gamma_t^*}.
\end{align}
{
Hence, the gradient with respect to $\gamma_t^*$ is given as
\begin{align}
\frac{\partial L_{\mathrm{A}}}{\partial \gamma_t^*}
&= 
\frac{\sum_{s}\rho_{t,s}\,
\big(e_{t,s}^* u_{t,s} - e_{t,s} u_{t,s}^*\big)}{2\gamma_t^*\sum_{t,s}\rho_{t,s}}
\\
&=
\frac{j\sum_{s}\rho_{t,s}\,
\Im \{e_{t,s}^* u_{t,s}\}}{\gamma_t^*\sum_{t,s}\rho_{t,s}}.
\end{align}
}

\end{appendices}

\bibliographystyle{IEEEtran}
\bibliography{ref}


 




\vfill

\end{document}